\documentclass[11pt]{article}
\usepackage[english]{babel}
\usepackage[utf8]{inputenc}
\usepackage[T1]{fontenc}
\usepackage{authblk}
\usepackage{graphicx}
\usepackage{csquotes}
\graphicspath{ {figures/} }
\usepackage{float}
\usepackage{tikz}
\tikzset{every picture/.style={line width=0.75pt}} %set default line width to 0.75pt
\usetikzlibrary{patterns}
\usepackage{amsmath,mathrsfs,dsfont}
\usepackage{amsthm}
\usepackage[all]{xy}
\usepackage{graphics}
\usepackage{dirtytalk}
\usepackage{tensor}
\usepackage{verbatim}
\usepackage{nicefrac}
\usepackage{amsfonts}
\usepackage{amssymb,latexsym}
\usepackage{color}
\usepackage{mathrsfs,dsfont}
\usepackage{physics}
\usepackage{mathtools}
\usepackage{bbm}
\usepackage{appendix}
\usepackage{slashed}
\usepackage{cite}
\usepackage{url}
\usepackage{hyperref}
\usepackage{comment}
\theoremstyle{definition}

\theoremstyle{remark}
\newtheorem{remark}{Remark}[section]

\numberwithin{equation}{section}

\newcommand{\nn}{\nonumber}
\newcommand{\bea}{\begin{eqnarray}}
\newcommand{\eea}{\end{eqnarray}}

\def\rlx{\relax\leavevmode}
\def\inbar{\vrule height1.5ex width.4pt depth0pt}
\def\IZ{\rlx\hbox{\sf Z\kern-.4em Z}}
\def\IN{\rlx\hbox{\rm I\kern-.18em N}}
\def\IO{\rlx\hbox{\,$\inbar\kern-.3em{\rm O}$}}
\def\IP{\rlx\hbox{\rm I\kern-.18em P}}
\def\IQ{\rlx\hbox{\,$\inbar\kern-.3em{\rm Q}$}}
\def\IF{\rlx\hbox{\rm I\kern-.18em F}}
\def\IG{\rlx\hbox{\,$\inbar\kern-.3em{\rm G}$}}
\def\IH{\rlx\hbox{\rm I\kern-.18em H}}
\def\II{\rlx\hbox{\rm I\kern-.18em I}}
\def\IK{\rlx\hbox{\rm I\kern-.18em K}}
\def\IL{\rlx\hbox{\rm I\kern-.18em L}}
\def\one{\hbox{{1}\kern-.25em\hbox{l}}}

\begin{document}

\title{Quantum Yang-Mills Charges in Strongly Coupled $2D$ Lattice QCD with Three Flavors}
\author[a]{Paulo A. Faria da Veiga\thanks{veiga@icmc.usp.br}}
\author[b]{Luiz A. Ferreira\thanks{laf@ifsc.usp.br}}
\author[b]{Henrique Malavazzi\thanks{henrique.malavazzi@usp.br}}
\author[a,b]{Ravi Mistry\thanks{ravi.mistry.r@gmail.com}}
\affil[a]{Instituto de Ciências Matemáticas e de Computação (ICMC), 
Universidade de São Paulo,
C.P. 668, 13560-970 São Carlos, SP, Brasil}
\affil[b]{Instituto de Física de São Carlos (IFSC), 
Universidade de São Paulo,
C.P. 369, 13560-970, São Carlos, SP, Brasil}

\date{\today}

\maketitle

\begin{abstract} 
We investigate the quantum properties of the truly gauge-invariant and conserved charges of two-dimensional Yang-Mills theories, focusing on lattice QCD in the  strong coupling regime. The construction of those charges uses the integral version of the $(1+1)$-dimensional Yang-Mills equations, and they correspond to the eigenvalues of a charge operator.  The gauge invariance of the charges  suggests that they are not confined, hence hadronic states may carry them. Using the path integral formalism with imaginary time (Euclidean), we evaluate the correlation functions of those charges on baryon and meson states through functional integrals over the gauge group $SU(N)$ ($ N=2,3$) and Grassmannian variables — the fermionic fields. Our results show that the expectation values of the lowest non-trivial charges are nonzero for baryon and meson states but vanish for non-gauge-invariant states, supporting the interpretation that hadrons indeed carry these charges. While renormalization effects and higher-order contributions remain to be analyzed, these findings point toward a potential link between gauge-invariant charges and confinement.
\end{abstract}

\maketitle

\tableofcontents

%%%%%%%%%%%%%%%%%%%%%%%%%%%%%%%%%%%%%%%%%%%%%%%%%%%%%%%%%%%%%%%%%%%%%%%
%%%%%%%%%%%%%%%%%%%%%%%%%%%%%%%%%%%%%%%%%%%%%%%%%%%%%%%%%%%%%%%%%%%%%%%
\section{\label{sec:level1}Introduction}
%%%%%%%%%%%%%%%%%%%%%%%%%%%%%%%%%%%%%%%%%%%%%%%%%%%%%%%%%%%%%%%%%%%%%%%
%%%%%%%%%%%%%%%%%%%%%%%%%%%%%%%%%%%%%%%%%%%%%%%%%%%%%%%%%%%%%%%%%%%%%%%
A remarkable fact about the laws of Physics governing the four fundamental interactions of Nature is that they are based on symmetry principles. The electromagnetic, the strong, and weak nuclear forces are based on the gauge principle, first formulated by H. Weyl \cite{weyl1929,dirac1931} for the case of Maxwell theory, and extended by C.N. Yang and R.L. Mills \cite{Yang:1954ek} (see also \cite{Weinberg:1996kr}) for the case of non-abelian gauge symmetries. A crucial requirement of such theories is that any observable must be invariant under gauge transformations. 

The equations of motion of gauge theories lead in a quite straightforward way to a number of conserved charges equal to the dimension of the gauge group. In the case of the electromagnetic interaction, the conserved charge is the electric charge, which is clearly gauge invariant. However, in the case of the non-abelian gauge theories, the conserved charges are not really gauge invariant. That is a long standing problem first noticed by Yang and Mills (see comment below eq. (16) of \cite{Yang:1954ek} or section 15.3 of \cite{Weinberg:1996kr}). Indeed, the Yang-Mills equation can be written as $j^{\mu}\equiv \partial_{\nu}F^{\nu\mu}=J^{\mu}-i\,e\,\left[A_{\nu},F^{\nu\mu}\right]$, where $A_{\mu}$ is the gauge potential, $F_{\mu\nu}$, the field tensor and $J_{\mu}$ the matter current. From the anti-symmetry of the field tensor, one has the conservation law $\partial_{\mu}j^{\mu}=0$, but the corresponding conserved charge $\int d^3x j_0$, is not gauge invariant. 

The construction of truly gauge invariant conserved charges for non-abelian gauge theories was first presented in \cite{Ferreira:2011ed,Ferreira:2012aj}, building on the developments in \cite{Alvarez:1997ma, Alvarez:2009dt}. The methods uses the integral form of the Yang-Mills equations, and is based on connection on loops spaces and the non-abelian Stokes theorem for two-form connections. The conserved charges are the eigenvalues of some operators that possess an isospectral time evolution, and there is an infinite number of charges. More recently, it was shown \cite{ferreira2025} that those charges Poisson commute and lead, on the sector of non-abelian electric and magnetic charges, to an exact integrability in non-abelian gauge theories. 

Since such conserved charges are gauge invariant, they are observable. In the case of QCD they are color singlet and so, in principle, are not confined. Therefore, the baryons and mesons could carry such charges. The calculation of the expectation values of those charges on hadron states in four dimensional Yang-Mills theories is a difficult task. For those reasons, we have constructed the equivalent of those charges in classical $(1+1)$-dimensional gauge theories in \cite{twoclassical}, with the objective of pursuing their role in the quantum $(1+1)$ Yang-Mills theory.  However, the quantum case is a largely uncharted territory, presenting many open questions that merit further investigation.

Understanding the quantum analog of the new charge operator is essential for addressing challenges related to gauge invariance and the interplay between classical and quantum symmetries. With this goal in mind, this work intends to take a step forward by laying the foundation for future investigations into the structure and behavior of quantum charge operators. A key related question is to understand what physical property this charge measures, in order to see how hadronic states, such as mesons and baryons, behave under these charge operators. How do the physical quantities behave under the action of this new gauge-invariant charge, and how does this charge contribute to our understanding of gauge symmetry in the quantum regime? Is there any relation between this quantum charge and confinement?

Of course, there are many treatments about confinement in the lattice QCD literature, specially in the fundamental paper \cite{fs}. In a mathematically rigorous approach confinement is manifest in the exponential decay of relevant correlations, since exponential decay implies a mass gap in the energy-momentum spectrum. In the present paper we are investigating the role of the new gauge invariant conserved charges, and searching for a possible relation among those charges and confinement. 

To achieve at least part of this goal, we consider a QCD-like simplified model with gauge groups ${\mathrm SU}(2)$ and ${\mathrm SU}(3)$, and with only three-flavor quark and antiquark fields. Namely, we take the flavors \enquote{$U$} (up), \enquote{$D$} (down), and \enquote{$S$} (strange). The analysis is performed in Euclidean spacetime dimensions on a square lattice with fixed lattice spacing $a > 0$. We adopt the  Wilson lattice action \cite{Gattringer:2010zz,Montvay:1994cy} and assume to be in the strong coupling regime, where the gauge coupling $g>0$ and the hopping parameter $\kappa>0$ satisfy the conditions $0 < \beta \equiv \beta(a)= (ag)^{-2}$ < $\kappa \ll 1$. 

In the strong coupling regime, the hopping term in the Wilson action dominates over the pure gauge plaquette action term. Consequently, for certain computations, the Wilson plaquette term can be neglected (see Eq. (\ref{total_act.}) below). We emphasize that the results obtained here can be generalized to higher Euclidean spacetime dimensions $d > 2$, to systems with a larger number of flavors, and gauge groups $SU(N\not= 2,3)$, with additional complexity and technical painstaking efforts.

We adopt the strong coupling approximation  because it simplifies the type of contributions to be considered when computing correlations (Green's functions) and, most important, it makes the evaluation of the gauge group integrals with the  Haar measure much simpler. In addition, we are interested in evaluating the expectation values of the charges on bound states, such as hadronic particle fields. It is worthwhile mentioning that the strong coupling regime has shown to be a good approximation. In previous papers,  by one of us, the existence of hadronic particles and some of their bound states was rigorously confirmed and quark confinement was verified up to some energy threshold in the energy-momentum spectrum \cite{BS1,BS2}. We also chose to work in $(1+1)$-dimensions since the spin is not really present, simplifying the algebraic computations. The strong coupling and the dimensionality are not mandatory. 

It is important to note that, in the strong coupling regime, the application of polymer expansions reveals significant properties. Starting with a finite lattice system in a box of volume $\Lambda$, for sufficiently small couplings \(\kappa\) and \(\beta\) (see \cite{Simon1993,Seiler:1982pw,FariadaVeiga:2008zz}), the thermodynamic limit $\Lambda\nearrow\infty$ of correlations exists, and truncated correlations exhibit exponential tree decay. The limiting correlations are invariant under lattice translations and extend to joint analytic functions of complex coupling parameters. Additionally, our model satisfies Osterwalder-Schrader positivity \cite{Osterwalder:1973dx,Osterwalder:1974tc,Glaser:1974hy, Glimm:1987ng,Seiler:1982pw}. This property ensures the existence of an underlying physical quantum mechanical Hilbert space \(\mathcal{H}\). Within this framework, by considering translations in time and spatial coordinates, we can define self-adjoint and self-commuting energy and momentum operators on \(\mathcal{H}\). Their joint spectrum is the system energy-momentum (E-M) spectrum \cite{Seiler:1982pw, Simon1993}, which can be accessed by analyzing the analyticity properties of a lattice Fourier transform of certain correlations. A relation between Hilbert space inner products and statistical mechanical correlation of the model is provided by a path integral (Feynman-Kac) type formula on the lattice \cite{Glimm:1987ng,FariadaVeiga:2008zz}.

In this context, for the gauge groups $G = {\mathrm SU}(N)$, with $N = 2,\; 3$, we define an Euclidean two-dimensional lattice quantum version of the YM current given in \cite{twoclassical}. We emphasize that, for two dimensions, the spin degrees of freedom are frozen and the spin indices $1$ and $2$ only refer to the particle-antiparticle distinction. Additionally, our results focus primarily on the case $N=2$, and we deal with $N=3$ only when our analysis incorporates baryon fields, which are not defined for $N=2$. 

The charge introduced in \cite{Alvarez:1997ma,Alvarez:2009dt} can be expanded in an infinite series of an arbitrary parameter, called $\lambda$, (see (\ref{chargeexpansion})), with each coefficient of the expansion associated with a conserved charge. However, in our analysis, we only consider the leading terms of the series, corresponding to $Q_M^{(1)}$ and $Q_M^{(2)}$ (see (\ref{chargebeta1}) and (\ref{chargebeta2b})), the first and second charges, respectively. Then, we define a quantum version of the classical charges and examine their quantum properties on the finite lattice when applied to composite meson and baryon fields. To this end, we study the corresponding path integral using the F-K like formula (\ref{F-K}), and compute the expectation values of the lattice charge operator for baryons and mesons (see (\ref{Expval O_f,g})). Although the focus is on the formula without explicitly using the partition function in the computation (i.e., just the numerator in (\ref{Expval O_f,g})), our computations capture the qualitative behaviour of the model. The discussion of the quantum current presented here will be scant, as our primary focus is on a lattice version of the classical current. Considering that the continuum limit $a \searrow 0$ of the model only exists after a renormalization procedure to remove ultraviolet singularities, a realistic quantum current and charge must incorporate corrections arising from renormalization and operator ordering effects. The analysis of the other charge terms, the higher orders of the expansion (\ref{chargeexpansion}), will eventually be the subject of further investigation

As a result, our analysis focused on the lattice version of $Q_M^{(2)}$, since we show that $Q_M^{(1)} = 0$ (see (\ref{chargebeta1})), for a semi-simple gauge group $G$. The charge $Q_{\ell}^{(2)}$ -- the implemented lattice version of $Q_M^{(2)}$ -- is zero for non-gauge invariant states, whereas it is nonzero for meson and baryon states.

Despite its limitations, our treatment captures many of the essential features necessary for exploring the interplay between gauge invariance and lattice discretization. We acknowledge that a fully rigorous treatment of the quantum current, incorporating these subtleties, remains an open challenge. Progress toward this goal will be reported in future work, where we aim to develop a more comprehensive framework for quantum currents on the lattice.

The paper is organized as follows. In Section \ref{sec:basics}, we outline the essential details of the underlying framework. Section \ref{sect. main} presents the main results of our study, while Section \ref{sect. conclusion} is dedicated to our conclusions. For completeness, we include several appendices that provide additional context related to some of the topics discussed and detailed computations.
%%%%%%%%%%%%%%%%%%%%%%%%%%%%%%%%%%%%%%%%%%%%%%%%%%%%%%%%%%%%%%%%%%%%%%%
%%%%%%%%%%%%%%%%%%%%%%%%%%%%%%%%%%%%%%%%%%%%%%%%%%%%%%%%%%%%%%%%%%%%%%%
%%%%%%%%%%%%%%%%%%%%%%%%%%%%%%%%%%%%%%%%%%%%%%%%%%%%%%%%%%%%%%%%%%%%%%%
\section{\label{sec:basics} Framework and Preliminaries}
%%%%%%%%%%%%%%%%%%%%%%%%%%%%%%%%%%%%%%%%%%%%%%%%%%%%%%%%%%%%%%%%%%%%%%%
%%%%%%%%%%%%%%%%%%%%%%%%%%%%%%%%%%%%%%%%%%%%%%%%%%%%%%%%%%%%%%%%%%%%%%%
%%%%%%%%%%%%%%%%%%%%%%%%%%%%%%%%%%%%%%%%%%%%%%%%%%%%%%%%%%%%%%%%%%%%%%%

\subsection{The classical charge operator and conserved charges}
 The conserved charges for classical  Yang-Mills theories in $(1+1)$-dimensions are discussed in \cite{twoclassical}. Here, we summarize the main aspects of the construction that serves as the backbone of our analysis on the lattice. 
 
The first ingredient of our construction is the so-called non-abelian Stokes theorem. Consider a $0$-form $B$, and a $1$-form $C_{\mu}$, $\mu=0,1$,  on the $(1+1)$-dimensional Minkowski space-time $\mathcal{M}$, with the metric $g_{\mu\nu}=\mathrm{diag}(1,-1)$, and both taking values on the Lie algebra $\mathfrak{g}$ of a Lie gauge group $G$. Let $\gamma(\sigma): [\sigma_i,\sigma_f]\longrightarrow x^{\mu}(\sigma) \in \mathcal{M}$ denote a path in $\mathcal{M}$, with $[\sigma_i, \sigma_f] \subset \mathbb{R}$, and define the holonomy of the connection $C_{\mu}$ on $\gamma$, through the ordinary differential equation  
 \bea \label{def. Wil.}
\frac{dW}{d\sigma}+C_{\mu}(x)\frac{dx^{\mu}(\sigma)}{d\sigma} W=0
\eea
By integrating (\ref{def. Wil.}), one obtains the path-ordered exponential, commonly referred to as the Wilson line
\bea \label{Wil. exp}
W(\sigma_i,\sigma)=\mathcal{P}\exp\left(-\int_{\sigma_i}^{\sigma}d\rho\,\, C_{\mu}(\rho)\frac{dx^{\mu}(\rho)}{d\rho}\right)W_R,
\eea
where $W_R\equiv W(\sigma_i,\sigma_i)$ stands for the integration constant. In particular, the series expansion of the Wilson line for the (totally) ordered partition of the subset $[\sigma_i,\sigma]$ where the upper bound $\sigma$, satisfies $\sigma \leq \sigma_f$, i.e.,
\bea
\sigma\geq\sigma_{0}\geq\sigma_{1}\geq\ldots\geq\sigma_{n-2}\geq\sigma_{i},
\eea
takes the form
\bea \label{Wilson p-ord.}
W(\sigma_i,\sigma)&=&\left(\mathbbm1-\int_{\sigma_{i}}^{\sigma}d\sigma_{0}C(\sigma_{0})+\int_{\sigma_{i}}^{\sigma}\int_{\sigma_{i}}^{\sigma_{0}}d\sigma_{0}d\sigma_{1}C(\sigma_{0})C(\sigma_{1})+\ldots+\nn
\right.\\
&& +\left. (-1)^{n-1}\,\int_{\sigma_{i}}^{\sigma}\int_{\sigma_{i}}^{\sigma_{0}}\ldots\int_{\sigma_{i}}^{\sigma_{n-2}}d\sigma_{0}d\sigma_{1}\ldots d\sigma_{n-1}C(\sigma_{0})C(\sigma_{1})\ldots C(\sigma_{n-1}) + \ldots\right)\,W_R
\nonumber
\eea
with $C(\sigma)=C_\mu(\sigma)\frac{dx^\mu(\sigma)}{d\sigma}$ and $n\geq 2$. The derivative of (\ref{Wilson p-ord.}) with respect to $\sigma$ straightforwardly leads to the expression for the integral curve associated with $W$ that passes through $W(\sigma_i)$. Since $W$ is matrix valued, by the uniqueness of the ODE solution (cf. Picard–Lindelöf theorem, see also Lipschitz continuity) one finds that $W$ is in fact $G$-valued, see e.g. \cite{Hamilton:2017gbn}, i.e., an element of the underlying group $G$.

We introduce a group element $V(\gamma)$ in $G$, through the exponential map,  as 
\bea\label{vdef}
V(\gamma)\equiv e^{W^{-1}(\gamma)\,B(x(\sigma_f))\,W(\gamma)}
\eea
The question we ask now is how $V(\gamma)$ changes when we change the end point of the curve $\gamma$ by an infinitesimal increment, i.e. $x^{\mu}(\sigma_f)\rightarrow x^{\mu}(\sigma_f+d\sigma)=x^{\mu}(\sigma_f)+\delta x^{\mu}$. Using a Baker-Campbell-Hausdorff type relation \cite{faraut,moore,sternberg}, we get that the variation of $V$ in terms of the argument of the exponential is \footnote{An alternative approach leads to a (proper) integral representation written 
in terms of the beta function, yielding the same result, see e.g.  \cite{Snider1964} and \cite{Wilcox:1967zz},
which demonstrates some mathematical methods to tackle the formula of interest.}
\bea\label{varv}
\delta V\,V^{-1}= \delta T+\frac{1}{2!}\, [T,\delta T]+\frac{1}{3!}\, [T,[T,\delta T]]+\ldots = \sum_{n=0}^{\infty}\frac{1}{(n+1)!}\, {\rm ad}^{n}_T\,\delta T
\eea
where $T\equiv W^{-1}(\gamma)\,B(x(\sigma_f))\,W(\gamma)$, and ${\rm ad}_T$ stands form the adjoint representation operator, i.e ${\rm ad}_T\,*\equiv [T,*]$. From (\ref{Wil. exp}), we have that
\bea
\delta W\, W^{-1}= -C_{\mu}(x)\frac{dx^{\mu}(\sigma)}{d\sigma}\,d\sigma
\eea
and so
\bea\label{vart}
\delta T= W^{-1}\, D_{\mu} B\,W\,\frac{dx^{\mu}(\sigma)}{d\sigma}\,d\sigma
\eea
with $D_{\mu}*\equiv \partial_{\mu}* + [C_{\mu},*]$. Therefore, from (\ref{varv}) and (\ref{vart}), we get a differential equation for $V$
\bea\label{neweqforv}
\frac{d\,V}{d\,\sigma}= W^{-1}\,L_{\mu}\,W\, \frac{dx^{\mu}(\sigma)}{d\sigma}\,V
\eea
with
\bea
L_{\mu}\equiv \sum_{j=0}^{\infty}\frac{1}{(j+1)!}\, {\rm ad}^{j}_B\, D_{\mu}B
\eea
Consequently, we can obtain $V$ by integrating (\ref{neweqforv}) along the curve $\gamma$, obtaining it as the holonomy of the connection $W^{-1}\,L_{\mu}\,W$. Since the result has to be the same as the one given in (\ref{vdef}), one obtains the non-abelian Stokes theorem
\bea\label{stokes}
 e^{W^{-1}(\gamma)\,B(x(\sigma_f))\,W(\gamma)}=\mathcal{P}\exp\left(\int_{\sigma_i}^{\sigma_f}d\sigma\,\, W^{-1}(\sigma)\,L_{\mu}(\sigma)\,W(\sigma)\frac{dx^{\mu}(\sigma)}{d\sigma}\right)V_R
 \eea
 with $V_R$ being an integration constant. The r.h.s. of (\ref{stokes}) can be written as a series of the form (\ref{Wilson p-ord.}), with $(-C(\sigma))$ replaced by $L(\sigma)\equiv W^{-1}(\sigma)\,L_{\mu}(\sigma)\,W(\sigma)\frac{dx^{\mu}(\sigma)}{d\sigma}$. Note that (\ref{stokes}) is a mathematical identity valid for any pair of differential forms $B$ and $C_{\mu}$. 
 
 The differential Yang-Mills equations in $(1+1)$-dimensions, for a gauge group $G$, are given by
 \bea\label{ymeqs}
 D_{\mu}F^{\mu\nu}=J^{\nu};\qquad\qquad\qquad \mu,\nu=0,1
 \eea
 where $J^{\mu}=J^{\mu}_a\,\tau_a$ is the matter current, $D_{\mu}=\partial_{\mu}*+i\,g[A_{\mu},*]$, is the covariant derivative, and $A_{\mu}=A_{\mu}^a\,\tau_a$, is the gauge field and
 \bea
 F_{\mu\nu}=\partial_{\mu}A_{\nu}-\partial_{\nu}A_{\mu}+i\,g\,[A_{\mu},A_{\nu}]
 \eea
 is the field tensor, or the curvature of the connection $A_{\mu}$. We have $\tau_a$, $a=1,2,\ldots{\rm dim}\,G$, is a basis for the Lie algebra of $G$, satisfying
 \bea
 [\tau_a,\tau_b]=i\,f_{abc}\,\tau_c\;;\qquad\qquad\qquad {\rm Tr}(\tau_a\,\tau_b)=\delta_{ab}
 \eea
 Introducing the Hodge duals
 \bea
 {\widetilde F}\equiv -\frac{1}{2}\,\varepsilon_{\mu\nu}\,F^{\mu\nu};\qquad \qquad{\widetilde J}_{\mu}\equiv - \varepsilon_{\mu\nu}\,J^{\nu};\qquad \qquad \varepsilon_{01}=1
 \eea
 we get that (\ref{ymeqs}) becomes
 \bea\label{ymeqsdual}
 D_{\mu}{\widetilde F}={\widetilde J}_{\mu}
 \eea
 
 The integral Yang-Mills equations are obtained from the Stokes theorem (\ref{stokes}), by using (\ref{ymeqsdual}) and making the replacements
 \bea\label{replace}
 C_{\mu}=i\,g\,A_{\mu}\;;\qquad\qquad\qquad B=i\,g\,\lambda\, {\widetilde F}
 \eea
 with $\lambda$ being an arbitrary (complex) parameter. Therefore, we get
 \bea\label{ymintegral}
 e^{i\,g\,\lambda\,W^{-1}(\gamma)\, {\widetilde F}(x(\sigma_f))\,W(\gamma)}=\mathcal{P}\exp\left(\int_{\sigma_i}^{\sigma_f}d\sigma\,\, W^{-1}(\sigma)\,{\cal K}_{\mu}(\sigma)\,W(\sigma)\frac{dx^{\mu}(\sigma)}{d\sigma}\right)V_R
 \eea
 with
 \bea\label{calkdef}
 {\cal K}_{\mu}= \sum_{n=0}^{\infty}\frac{(i\,g\,\lambda)^{n+1}}{(n+1)!}\, {\rm ad}^{n}_{\widetilde F}\, {\widetilde J}_{\mu}
 \eea
The Wilson line $W$ is now obtained from  (\ref{def. Wil.}), and the replacements (\ref{replace}), and so
  \bea \label{wilson}
\frac{dW}{d\sigma}+i\,g\,A_{\mu}(x)\frac{dx^{\mu}(\sigma)}{d\sigma} W=0
\eea
 By considering an infinitesimal path, where $\sigma_f\rightarrow \sigma_i$, we see that the integration constant in (\ref{ymintegral}) must be
 \bea\label{intconstant}
 V_R=e^{i\,g\,\lambda\,W^{-1}_R\, {\widetilde F}(x(\sigma_i))\,W_R}
 \eea
 with $W_R$ being the integration constant appearing in (\ref{Wil. exp}). 
 
The integral equation  (\ref{ymintegral}) is equivalent to the differential Yang-Mills equations (\ref{ymeqsdual}) (or equivalently (\ref{ymeqs})). Indeed, it was obtained from the (identity) Stokes theorem (\ref{stokes}) and the Yang-Mills equations  (\ref{ymeqsdual}). Alternatively, by considering an infinitesimal curve $\gamma$, and Taylor expanding both sides of (\ref{ymintegral}) in a power series in the length of the path, one gets the Yang-Mills equation (\ref{ymeqsdual}) in lowest order. 
 
 Under a local gauge transformation 
 \bea
 A_{\mu}(x)\rightarrow \varrho(x)\,A_{\mu}(x)\,\varrho^{-1}(x)+\frac{i}{g}\partial_{\mu}\varrho(x)\,\varrho^{-1}(x)\;;\qquad\qquad \varrho\in G
 \eea
 we have that
 \bea
 F_{\mu\nu}(x)\rightarrow \varrho(x)\,F_{\mu\nu}(x)\,\varrho^{-1}(x)\;;\qquad J_{\mu}(x)\rightarrow \varrho(x)\,J_{\mu}(x)\,\varrho^{-1}(x)
  \eea
  and
  \bea 
 W(\sigma)\rightarrow \varrho(\sigma)\,W(\sigma)\,{\widetilde \varrho}^{-1}_R\;; \qquad\qquad 
{\widetilde \varrho}_R\equiv  W_R^{-1}\,\varrho(\sigma_i)\,W_R
\eea
Therefore,
\bea\label{gaugetransf}
W^{-1}(\gamma)\, {\widetilde F}(x(\sigma_f))\,W(\gamma)&\rightarrow& {\widetilde \varrho}_R\,W^{-1}(\gamma)\, {\widetilde F}(x(\sigma_f))\,W(\gamma)\,{\widetilde \varrho}^{-1}_R
\\
W^{-1}(\sigma)\,{\cal K}_{\mu}(\sigma)\,W(\sigma)&\rightarrow& {\widetilde \varrho}_R\, W^{-1}(\sigma)\,{\cal K}_{\mu}(\sigma)\,W(\sigma)\,{\widetilde \varrho}^{-1}_R
\\
V_R&\rightarrow& {\widetilde \varrho}_R\,V_R\,{\widetilde \varrho}^{-1}_R
\eea
Consequently, both sides of the integral Yang-Mills equation (\ref{ymintegral}) transform under conjugation by the constant group element ${\widetilde \varrho}_R$, and so are covariant under local gauge transformations.

The important property following from the integral equation   (\ref{ymintegral}) is that the eigenvalues of the operators on both sides of  (\ref{ymintegral}) are independent of the path $\gamma$, as long as their end points are kept fixed. In order to see that, consider a curve $\gamma^{\prime}$ with the same endpoints as $\gamma$. Therefore, $\gamma\circ{\gamma^{\prime}}^{-1}$ is a closed path. We can then use the non-abelian Stokes theorem for a $1$-form connection \cite{Alvarez:1997ma,Ferreira:2012aj,Ferreira:2011ed} to write
\bea
W({\gamma^{\prime}}^{-1})\,W(\gamma)={\cal P} \exp\left(i\,g\, \int_{\Sigma} d\sigma\,d \tau\, W^{-1}\, F_{\mu\nu}\,W\,\frac{d\,x^{\mu}}{d\,\sigma}\,\frac{d\,x^{\nu}}{d\,\tau}\right)\equiv H(\Sigma)
\eea 
where $\Sigma$ is any  surface with border $\partial \Sigma= \gamma\circ{\gamma^{\prime}}^{-1}$. The surface $\Sigma$ is scanned with closed loops, labelled by $\tau$, based on a reference point on its border, and each loop is parameterized by $\sigma$. It then follows that
\bea
e^{i\,g\,\lambda\,W^{-1}(\gamma^{\prime})\, {\widetilde F}(x(\sigma_f))\,W(\gamma^{\prime})}=H(\Sigma)\,e^{i\,g\,\lambda\,W^{-1}(\gamma)\, {\widetilde F}(x(\sigma_f))\,W(\gamma)}\,H(\Sigma)^{-1}
\eea
So, if we change the path $\gamma$ to another path $\gamma^{\prime}$, with the same endpoints, the operator on the l.h.s. of (\ref{ymintegral}) changes by a conjugation by $H(\Sigma)$. This means that its eigenvalues are independent of the path, as long as its endpoints are kept fixed. Hence (\ref{ymintegral}) implies that the operator on its r.h.s. has the same property. Since the eigenvalues of a matrix can be written functionally in terms of traces of its powers, we have that the quantities
\bea\label{pathindeigen}
{\rm Tr}\left(e^{i\,g\,\lambda\,W^{-1}(\gamma)\, {\widetilde F}(x(\sigma_f))\,W(\gamma)}\right)^M={\rm Tr}\left(\mathcal{P}\exp\left(\int_{\sigma_i}^{\sigma_f}d\sigma\,\, W^{-1}(\sigma)\,{\cal K}_{\mu}(\sigma)\,W(\sigma)\frac{dx^{\mu}(\sigma)}{d\sigma}\right)V_R\right)^M
\eea
 are path independent, where $M$ is an integer, indicating powers of the operator. Such property can be used to construct conserved quantities as we now explain. 
 
 Let us consider two paths with the same endpoints. The first one, denoted by $\Gamma_t$, is a straight line parallel to the $x^1$-axis, at a fixed time $t=t$, and going from $x^1=-\infty$ to $x^1=+\infty$. The second one is made of three parts, $\Gamma_{-\infty}^{-1}\circ\Gamma_0\circ\Gamma_{+\infty}$, where $\Gamma_0$ is a straight line, parallel to the $x^1$-axis, at fixed time $t=0$, going from $x^1=-\infty$ to $x^1=+\infty$. The paths $\Gamma_{\pm \infty}$ are straight lines, parallel to the $x^0$-axis, positioned at $x=\pm \infty$, and going from $t=0$ to $t=t$. The operator on the r.h.s of (\ref{ymintegral}) is obtained by integrating (\ref{neweqforv}), with $L_{\mu}$ replaced by ${\cal K}_{\mu}$, given in (\ref{calkdef}). Therefore, it is the holonomy of the connection $W^{-1}\,{\cal K}_{\mu}\,W$, and so it satisfies the usual rules of decomposition under sectioning of paths. Indeed, if $\Gamma= \Gamma_1\circ\Gamma_2$, we have from (\ref{neweqforv}), with $L_{\mu}$ replaced by ${\cal K}_{\mu}$, that  
 \bea
 V(\Gamma)=V(\Gamma_2)\,V(\Gamma_1)
 \eea
 with 
\bea
V(\Gamma_a)\equiv \mathcal{P}\exp\left(\int_{\Gamma_a}d\sigma\,\, W^{-1}(\sigma)\,{\cal K}_{\mu}(\sigma)\,W(\sigma)\frac{dx^{\mu}(\sigma)}{d\sigma}\right)
\eea
with no integration constant. However, we have to take into account that $W$ in $V(\Gamma_2)$ is calculated from the initial point of $\Gamma$, which is the same initial point of $\Gamma_1$. Therefore, if we denote by ${\widetilde V}(\Gamma_2)$, the holonomy obtained by integrating (\ref{neweqforv}), with $L_{\mu}$ repalced by ${\cal K}_{\mu}$, but with $W$ calculated from the initial point of $\Gamma_2$, with integration constant for $W$ set to unity, we have that
\bea
 V(\Gamma)=V(\Gamma_2)\,V(\Gamma_1)=W^{-1}(\Gamma_1)\,{\widetilde V}(\Gamma_2)\,W(\Gamma_1)\,V(\Gamma_1)
 \eea
 We take the initial point of paths $\Gamma_t$ and $\Gamma_{-\infty}^{-1}\circ\Gamma_0\circ\Gamma_{+\infty}$, to be $x_R\equiv(x^0=c\,t,x^1=-\infty)$. Note that the initial point of $\Gamma_{-\infty}^{-1}$ is also $x_R$. Therefore, from the path independence of the quantities given in (\ref{pathindeigen}), we have that
 \bea
{\rm Tr}\left(V(\Gamma_t)\,V_R\right)^M&=&{\rm Tr}\left( V(\Gamma_{+\infty})\,V(\Gamma_0)\,V(\Gamma_{-\infty}^{-1})\,V_R\right)^M
\nonumber\\
&=&{\rm Tr}\left( V(\Gamma_{+\infty})\,W^{-1}(\Gamma_{-\infty}^{-1})\,{\widetilde V}(\Gamma_0)\,W(\Gamma_{-\infty}^{-1})\,V(\Gamma_{-\infty}^{-1})\,V_R\right)^M
\\
&=&{\rm Tr}\left(W^{-1}(\Gamma_{-\infty}^{-1})\,W^{-1}(\Gamma_0)\, {\widetilde V}(\Gamma_{+\infty})\,W(\Gamma_0)\,{\widetilde V}(\Gamma_0)\,W(\Gamma_{-\infty}^{-1})\,V(\Gamma_{-\infty}^{-1})\,V_R\right)^M
\nonumber
\eea
On the paths $\Gamma_t$ and $\Gamma_0$, we have $x^0=t$ and $x^0=0$, respectively, and $x^1=\sigma$. On the paths $\Gamma_{\pm \infty}$, we have $x^0=\sigma$, and $x^1=\pm \infty$. Therefore, on the paths $\Gamma_{\pm \infty}$, we have the path ordered integral of $W^{-1}\,{\cal K}_{1}(\sigma)\,W$, which is linear in $J_1$, and $W(\Gamma_{\pm \infty})$ only involves $A_0$. Consequently, if we impose the boundary conditions 
\bea\label{j1boundcond}
J_1\rightarrow \frac{1}{{(x^1})^{\delta}} \qquad {\rm as}\qquad x^1\rightarrow \pm \infty\qquad {\rm with} \qquad \delta>0
\eea
and 
\bea\label{a0boundcond}
A_0\rightarrow \frac{1}{{(x^1})^{\delta^{\prime}}} \qquad {\rm as}\qquad x^1\rightarrow - \infty\qquad {\rm with} \qquad \delta^{\prime}>0
\eea
we get that
\bea
V(\Gamma_{-\infty})\rightarrow \one\;;\qquad \qquad{\widetilde V}(\Gamma_{+\infty})\rightarrow \one\;;\qquad \qquad
W(\Gamma^{-1}_{-\infty})\rightarrow W_R
\eea
Note that, from (\ref{ymeqsdual}), (\ref{j1boundcond}) and (\ref{a0boundcond}) 
\bea\label{ftildeconstant}
\partial_0 {\widetilde F}\rightarrow 0 \qquad\qquad {\rm as}\qquad\qquad x^1\rightarrow - \infty
\eea
Therefore, we get that
\bea
{\rm Tr}\left(V(\Gamma_t)\,V_R\right)^M={\rm Tr}\left({\widetilde V}(\Gamma_0)\,W_R\,V_R\,W_R^{-1}\right)^M={\rm Tr}\left({\widetilde V}(\Gamma_0)\,e^{i\,g\,\lambda\, {\widetilde F}(x^1=-\infty)}\right)^M
\eea
where ${\widetilde F}(x^1=-\infty)$ is the value of ${\widetilde F}$ at $x^1=-\infty$, for any value of time. Indeed, from (\ref{ftildeconstant}), we have that ${\widetilde F}$ is constant in time at $x^1=-\infty$. 

As we have shown above, ${\widetilde V}(\Gamma_0)$ is calculated with the integration constant for $W$, at the initial point of $\Gamma_0$, set to unity. The initial point of $\Gamma_0$ is $x_{R_0}\equiv(x^0=0,x^1=-\infty)$. If instead, we set the value $W$ at $x_{R_0}$ to be $W_{R_0}$, we get that the holonomy $V$ on $\Gamma_0$, becomes ${\widehat V}(\Gamma_0)=W_{R_0}^{-1}\,{\widetilde V}(\Gamma_0)\,W_{R_0}$. Therefore, we get
\bea
{\rm Tr}\left(V(\Gamma_t)\,V_R\right)^M={\rm Tr}\left({\widehat V}(\Gamma_0)\,V_{R_0}\right)^M\;;\qquad\qquad
V_{R_0}\equiv e^{i\,g\,\lambda\, W_{R_0}^{-1}\,{\widetilde F}(x^1=-\infty)\,W_{R_0}}
\eea

Consequently the charges
\bea
Q_M(\lambda)\equiv \frac{1}{M}\,{\rm Tr}\left(V(\Gamma_t)\,V_R\right)^M\;;\qquad \qquad {\rm with} \qquad \qquad
V_R=e^{i\,g\,\lambda\,W^{-1}_R\, {\widetilde F}(x^1=-\infty)\,W_R}
\eea
are conserved in time. From (\ref{gaugetransf}), we observe that such charges are indeed gauge invariant.

By expanding the charge operator in a power series in $\lambda$, we get that each term of the series is a conserved quantity, and so we obtain an infinity of conserved charges $Q_M^{(j)}$, i.e.
\bea\label{chargeexpansion}
Q_M(\lambda)= \sum_{j=0}^{\infty} \lambda^j\, Q_M^{(j)}
 \eea
The first charge in order $\lambda^0$ is trivial, i.e.  $Q_M^{(0)}=\frac{1}{M}\,{\rm Tr}\one^M$. The charge in the first order is
\bea\label{chargebeta1}
Q_M^{(1)}=i\,g\,{\rm Tr}\left( \int_{-\infty}^{\infty}dx^1\, W^{-1}(x^1)\,J_0(x^1)\,W(x^1)+ W_R^{-1}\,{\widetilde F}(x^1=-\infty)\,W_R \right)
\eea
For a simple Lie algebra, any generator can be expressed as the commutator of two others, and so, the trace of any generator vanishes in any representation. Therefore, (\ref{chargebeta1}) vanishes if the gauge group $G$ is simple. 

The charge in second order ($\lambda^2$) is
\bea\label{chargebeta2}
Q_M^{(2)}&=&(i\,g)^2\,{\rm Tr}\left( \int_{-\infty}^{\infty}dx^1\, W^{-1}(x^1)\,J_0(x^1)\,W(x^1)\,\int_{-\infty}^{x^1}d{\bar x}^1\, W^{-1}({\bar x}^1)\,J_0({\bar x}^1)\,W({\bar x}^1)
\right.
\nonumber\\
&+&\left. \frac{(M-1)}{2}\, \left[\int_{-\infty}^{\infty}dx^1\, W^{-1}(x^1)\,J_0(x^1)\,W(x^1)\right]^2+ \frac{(M-1)}{4}\, \left[W_R^{-1}\,{\widetilde F}(x^1=-\infty)\,W_R\right]^2
\right. 
\nonumber\\
&+& \left. M
\, W_R^{-1}\,{\widetilde F}(x^1=-\infty)\,W_R\,\int_{-\infty}^{\infty}dx^1\, W^{-1}(x^1)\,J_0(x^1)\,W(x^1)\right)
\eea
Using the cyclic property of the trace, we can write
\bea
&&{\rm Tr}\left( \int_{-\infty}^{\infty}dx^1\, W^{-1}(x^1)\,J_0(x^1)\,W(x^1)\,\int_{-\infty}^{x^1}d{\bar x}^1\, W^{-1}({\bar x}^1)\,J_0({\bar x}^1)\,W({\bar x}^1)\right)
\nonumber\\
&&=\frac{1}{2}\, {\rm Tr}\left[\int_{-\infty}^{\infty}dx^1\, W^{-1}(x^1)\,J_0(x^1)\,W(x^1)\right]^2
\eea
In our applications in this paper, we shall work with the boundary condition ${\widetilde F}(x^1=-\infty)=0$. Therefore, the charge in the second order we shall use is given by 
\bea\label{chargebeta2b}
Q_M^{(2)}=\frac{M}{2}\,(i\,g)^2\,{\rm Tr} \left[\int_{-\infty}^{\infty}dx^1\, W^{-1}(x^1)\,J_0(x^1)\,W(x^1)\right]^2.
\eea

%%%%%%%%%%%%%%%%%%%%%%%%%%%%%%%%%%%%%%%%%%%%%%%%%%%%%%%%%%%%%%%%
%%%%%%%%%%%%%%%%%%%%%%%%%%%%%%%%%%%%%%%%%%%%%%%%%%%%%%%%%%%%%%%%
%%%%%%%%%%%%%%%%%%%%%%%%%%%%%%%%%%%%%%%%%%%%%%%%%%%%%%%%%%%%%%%%
\subsection{1+1 quantum lattice Wilson action: pure gauge and matter}
%%%%%%%%%%%%%%%%%%%%%%%%%%%%%%%%%%%%%%%%%%%%%%%%%%%%%%%%%%%%%%%%
%%%%%%%%%%%%%%%%%%%%%%%%%%%%%%%%%%%%%%%%%%%%%%%%%%%%%%%%%%%%%%%%
%%%%%%%%%%%%%%%%%%%%%%%%%%%%%%%%%%%%%%%%%%%%%%%%%%%%%%%%%%%%%%%%

We consider a two-dimensional Euclidean spacetime model comprising both gauge and matter components. Specifically, our focus is on a lattice action featuring two flavors, $U$ and $D$; as well as three flavours, $U$, $D$, and $S$, in the strong coupling regime, i.e. $0<g^{-2}\ll\kappa$. 

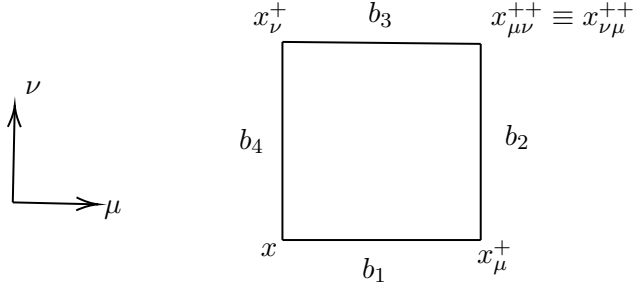
\begin{figure}
    \centering
    \begin{tikzpicture}[x=0.75pt,y=0.75pt,yscale=-1,xscale=1]

%Straight Lines [id:da6773873485647783] 
\draw    (306.33,89) -- (306.33,189) ;
%Straight Lines [id:da6514860200700145] 
\draw    (306.33,189) -- (406.33,189) ;
%Straight Lines [id:da16707029666029127] 
\draw    (306.33,89) -- (406.33,90) ;
%Straight Lines [id:da5041824733289477] 
\draw    (406.33,90) -- (406.33,189) ;
%Straight Lines [id:da16919828634467637] 
\draw    (170.33,170) -- (171.29,123) ;
\draw [shift={(171.33,121)}, rotate = 91.17] [color={rgb, 255:red, 0; green, 0; blue, 0 }  ][line width=0.75]    (10.93,-3.29) .. controls (6.95,-1.4) and (3.31,-0.3) .. (0,0) .. controls (3.31,0.3) and (6.95,1.4) .. (10.93,3.29)   ;
%Straight Lines [id:da7562544375937372] 
\draw    (170.33,170) -- (211.33,170.95) ;
\draw [shift={(213.33,171)}, rotate = 181.33] [color={rgb, 255:red, 0; green, 0; blue, 0 }  ][line width=0.75]    (10.93,-3.29) .. controls (6.95,-1.4) and (3.31,-0.3) .. (0,0) .. controls (3.31,0.3) and (6.95,1.4) .. (10.93,3.29)   ;

% Text Node
\draw (294,189.4) node [anchor=north west][inner sep=0.75pt]    {$x$};
% Text Node
\draw (403,187.4) node [anchor=north west][inner sep=0.75pt]    {$x_{\mu }^{+}$};
% Text Node
\draw (410,67.4) node [anchor=north west][inner sep=0.75pt]    {$x_{\mu \nu }^{++} \equiv x_{\nu \mu }^{++}$};
% Text Node
\draw (290,67.4) node [anchor=north west][inner sep=0.75pt]    {$x_{\nu }^{+}$};
% Text Node
\draw (175,107.4) node [anchor=north west][inner sep=0.75pt]    {$\nu $};
% Text Node
\draw (215,167.4) node [anchor=north west][inner sep=0.75pt]    {$\mu $};
% Text Node
\draw (345,196.4) node [anchor=north west][inner sep=0.75pt]    {$b_{1}$};
% Text Node
\draw (417,129.4) node [anchor=north west][inner sep=0.75pt]    {$b_{2}$};
% Text Node
\draw (348,67.4) node [anchor=north west][inner sep=0.75pt]    {$b_{3}$};
% Text Node
\draw (283,131.4) node [anchor=north west][inner sep=0.75pt]    {$b_{4}$};

\end{tikzpicture}

    \caption{The plaquette: a depiction of bonds connecting the nearest neighbour points.}
    \label{Fig. plaq.}
\end{figure}

Let $\Lambda \subset a(\mathbb{Z} + 1/2)\times a\mathbb{Z} \subset \mathbb{R}^2$ be a finite lattice with lattice spacing $a > 0$. Furthermore, the lattice $\Lambda$ is assumed to consist of $L \in \mathbb{N}$ sites, where $L$ is taken to be even for technical convenience. A site, an arbitrary lattice point, will be denoted by $x \equiv (x_0, x_1) \in \Lambda$, and we let the time direction be denoted by $0$. The fermionic local fields are considered as \emph{independent} Grassmann variables \cite{Gattringer:2010zz}, $\psi(x)$ (quarks) and $\bar{\psi}(x)$ (antiquarks), attached to each lattice site. As above, the underlying gauge group $G$ is the compact Lie group $SU(N)$ of dimension $N^2 - 1$. A lattice bond connecting nearest-neighbor points is written as
\bea \label{bond eq.}
b=(x,x+ae_\mu),\quad\mathrm{with}\quad e_\mu\,\text{-- unit vector},\quad\mu=0,1.
\eea
Henceforth, we will use shorthand notations for the translation of lattice points; for example, $x^\pm_\mu \equiv x \pm a e_\mu$, $x^{++}_{\mu\nu} \equiv x^+_\mu + a e_\nu$, and so on. For each bond $b$, there is a corresponding gauge group element $U_b \in G$, referred to as a \emph{bond variable}. If $b$ is the bond specified in equation (\ref{bond eq.}), we denote it by $U_b \equiv g_{x x^+_\mu}$, where its components are $(g_{x x^+_\mu})_{ab}$, with $a, b = 1, 2, \ldots, N$ being the {\em color} indices, and $N$ the dimension of the defining representation \cite{Hamilton:2017gbn}. A \emph{plaquette} is a minimal lattice square and carries the \emph{plaquette variable} $U_p$, defined as the ordered product of lattice bond variables (see Fig. \ref{Fig. plaq.}), i.e.,
\bea \label{p-term 1}
U_{p}\doteq U_{b_{1}}U_{b_{2}}U_{b_{3}}^{\dagger}U_{b_{4}}^{\dagger}.
\eea
Here, the bonds are oriented according to the following prescription: if $b$ is in the positive $\mu$ or $\nu$ direction, we write $U_b$; otherwise, we write $U_b^\dagger$. For the latter case, we have $U_b^\dagger \equiv U_b^{-1}$, since $U_b \in G$ is unitary. To facilitate later use, we rewrite equation (\ref{p-term 1}) in a more convenient form as
\bea \label{p-term 2}
U_p \doteq g_{xx_\mu^+} g_{x_\mu^+x_{\mu\nu}^{++}}g^{-1}_{x_{\mu\nu}^{++}x_\nu^+}g^{-1}_{x_\nu^+x}.
\eea

In our formulation, we consider the gamma spin matrices $\{\gamma_0, \gamma_1\}$ as generators of the relevant Clifford algebra, which is isomorphic to $M_2(\mathbb{R})$. These gamma matrices are taken to be Hermitian, and identified with the real Pauli matrices:
\bea
\gamma_0 = \begin{pmatrix} 1 & 0 \\ 0 & -1 \end{pmatrix}, \quad \text{and} \quad \gamma_1 = \begin{pmatrix} 0 & 1 \\ 1 & 0 \end{pmatrix},
\eea
which satisfy the anticommutation relations:
\bea
\{\gamma_\mu, \gamma_\nu\} = 2 \delta_{\mu\nu} \mathbbm{1}_2, \quad \text{and} \quad \mathrm{tr}(\gamma_\mu) = 0,
\eea
where $\mathbbm{1}_2$ is the $2 \times 2$ identity matrix. These matrices form the basis for the underlying Clifford algebra and play a fundamental role in the description of spinor fields in two-dimensional spacetime. As mentioned before, the two degrees of freedom in the spinors, since we are in dimension two and the spins are frozen, only refer to particle and antiparticle states, cf. remark \ref{+-}.

We now define other useful matrices in terms of $\gamma_0$ and $\gamma_1$, which will be instrumental in writing the action for the system under consideration. Let $\epsilon,\epsilon^\prime = \pm 1$, then
\bea
\Gamma^{\epsilon e_\mu}\doteq -\mathbbm{1}_{2}\pm\gamma_{\mu},
\eea
%\vspace{-0.5cm}
satisfies the following properties \cite{FariadaVeiga:2004rf}:
\begin{enumerate}
    \item $\Gamma^{\epsilon e_{\mu}}\Gamma^{-\epsilon e_{\mu}}=0$,
    \item $\Gamma^{\epsilon e_{\mu}}\Gamma^{\epsilon e_{\mu}}= -2\Gamma^{\epsilon e_{\mu}}$,
    \item $\Gamma^{\epsilon e_{\mu}}\Gamma^{\epsilon^{\prime}e_{\nu}}=2\mathbbm{1}_{2}-\Gamma^{-\epsilon^{\prime}e_{\nu}}\Gamma^{-\epsilon e_{\mu}}$.
\end{enumerate}

Let $\mu=0,1$. Let $a$, $x$, $\Lambda$, $\epsilon$, $g_{xx^{\pm}_{\mu}}$, $U_p$ and $\Gamma^{\epsilon e_\mu}$ be as above. Then the total action in two-dimensional Euclidean spacetime is given by \cite{Gattringer:2010zz,Montvay:1994cy}
\bea\label{total_act.}
S_{\Lambda}(\psi,\bar{\psi},U_{p})=S_g(U_p)+S_f(\psi,\bar{\psi}),
\eea
where the quantities on the right-hand side are defined as follows. The pure gauge part (the trace is taken over the group $G$) 
\bea \label{act. g}
S_g(U_p)\doteq\frac{1}{(ag)^2}\sum_{p\in\Lambda}A_{p}(U_{p})\qquad\mathrm{with}\qquad A(U_{p})\doteq2\,\mathrm{Re}\,\,\mathrm{Tr}\left(\mathbbm{1}-U_{p}\right).
\eea
Note that using the Hilbert-Schmidt (HS) norm $\| \mathcal O \|_{HS}=\left[{\rm Tr} (\mathcal O^\dagger \mathcal O)\right]^{1/2}$, where $\mathcal{O}$ is a finite dimensional matrix and $\dagger$ denotes the usual conjugate transpose, one can show that
\bea \label{HS norm}
A(U_{p})=2\,\mathrm{Re}\,\,\mathrm{Tr}\left(\mathbbm{1}-U_{p}\right)=||U_{p}-\mathbbm{1}||_{\mathrm{HS}}^{2},
\eea
such that $A(U_{p})$ is nonnegative. The (Fermionic) matter part of the action is
\bea \label{act. f}
S_f(\psi,\bar{\psi},U_b)\doteq\frac{\kappa a}{2}\sum_{x,\epsilon,\mu}\bar{\psi}_{a\alpha f}(x)\Gamma_{\alpha\beta}^{\epsilon e_{\mu}}\left(g_{xx_{\mu}^{\epsilon}}\right)_{ab}\psi_{b\beta f}\left(x_{\mu}^{\epsilon}\right)+a^{2}\sum_{x\in\Lambda}\bar{\psi}_{a\alpha f}(x)M_{\alpha\beta}\psi_{a\beta f}(x),
\eea
where we are summing over repeated indices,
\bea\label{mass_term}
M_{\alpha\beta}=\left(m+\frac{2\kappa}{a}\right)\delta_{\alpha\beta}, \qquad\alpha,\beta=1,2,
\eea
with $\kappa>0$ denoting the hopping parameter and $m>0$ the bare mass. Here, $\psi_{b\beta f}$ and ${\bar \psi}_{b\beta f}$ are Wilson fermions (no doubling), with $b=1,2$, labelling the color $SU(2)$ doublet representation components, $\beta$ is the spin state index (particle and anti-particle), and $f$ being the flavor (isospin) label. In addition, $\psi$ and $\bar{\psi}$, treated as independent Grassmann variables, are nilpotents of degree 2, a consequence of the anti-commutation relations. One easily checks that the formal continuum limit $a \searrow 0$ recovers the usual smooth continuous field action.

In the strong coupling regime, the Osterwalder-Schrader-Seiler positivity holds (see \cite{osterwalder1978gauge, Osterwalder:1974tc, Osterwalder:1973dx}), and there is an underlying quantum mechanical Hilbert space $\mathcal{H}$ associated with our lattice model.
%%%%%%%%%%%%%%%%%%%%%%%%%%%%%%%%%%%%%%%%%%%%%%%%%%%%%%%%%%%%%%%%%%%
%%%%%%%%%%%%%%%%%%%%%%%%%%%%%%%%%%%%%%%%%%%%%%%%%%%%%%%%%%%%%%%%%%%
\subsection{Path integral: expectation values on the lattice}
%%%%%%%%%%%%%%%%%%%%%%%%%%%%%%%%%%%%%%%%%%%%%%%%%%%%%%%%%%%%%%%%%%%
%%%%%%%%%%%%%%%%%%%%%%%%%%%%%%%%%%%%%%%%%%%%%%%%%%%%%%%%%%%%%%%%%%%
Let $\mathcal{O}(\psi,\bar{\psi},g)$, henceforth $\mathcal{O}$, be an observable. Then the inner product on $\mathcal{H}$, for $F$ and $G$ restricted to $x_0=1/2$, is related to the Feynman-Kac (F-K) like formula (cf. thm. X. 68 in \cite{RS1975ii}, see also \cite{Wipf:2013vp,Glimm:1987ng,Seiler:1982pw}) as follows. 
\bea \label{F-K}
\left(\Theta G,\mathcal{O}F\right)_{\mathcal{H}}=\expval{\mathcal{O}FG}_{F-K}\quad\forall\,F,\,G\in \mathcal{H},
\eea
where $\Theta$ denotes an appropriate antilinear operator on $\mathcal{H}$ that implements time reflection and acts on the fields $\psi$ and $\bar{\psi}$ in the following manner. For $x=(x_0,x_1)$, we have
\bea \label{Theta act.}
\Theta \psi(x)=\bar{\psi}(Tx)\gamma_0\quad\mathrm{and}\quad\Theta\bar{\psi}(x)=\gamma_0\psi(Tx),
\eea
with $Tx\doteq (-x_0,x_1)$. Moreover, $\Theta$ satisfies the following properties \cite{Seiler:1982pw}
\begin{enumerate}
    \item $\Theta(FG)=\Theta(G)\Theta(F)\quad\forall\,F,\,G\in \mathcal{H}$,
    \item $\Theta$ is an involution on $\mathcal{H}$, i.e., $\Theta^2=\mathbbm{1}_\mathcal{H}$.
\end{enumerate}
Note that the second property is obtained as a result of applying the first one and (\ref{Theta act.}) on the product of arbitrary state vectors.

Before we outline the underlying framework used in our computation, we draw attention to the following remark that might be useful for a rigorous analysis of the problem.
\begin{remark}
The particle spectrum was analyzed in Refs. \cite{Klu,KS} at an early stage of lattice QCD, in the strong coupling approximation. Some works by one of us and O'Carroll extends these analyses to a rigorous level Using a lattice with fixed finite spacing, the baryonic sector of the energy-momentum spectrum in strongly coupled lattice QCD. In the 2004 study \cite{FariadaVeiga:2004rf} (in collaboration with Schor), baryons and anti-baryons are obtained as bound states of three (anti-)quarks, with asymptotic masses \(M_s = -3 \ln \kappa + r_s(\kappa)\) and small mass splittings observable in \(d=2\) and \(d=3\), where \(d\) denotes the spacetime dimensionality of the lattice. The term $r_s(\kappa)$ accounts for additional factors or refinements in the mass that go beyond the simple logarithmic dependence on $\kappa$, and it likely contains higher-order corrections that become relevant at specific values of $\kappa$, for small $\kappa$. While the construction of the physical Hilbert space \(\mathcal{H}\) using gauge-invariant correlation functions requires a careful treatment \cite{Seiler:1982pw}, the 2004 study demonstrates that the baryonic spectrum appears as isolated energy-momentum dispersion curves within the subspace \(\mathcal{H}_o\), generated by states with an odd number of fermions. The 2008 paper \cite{FariadaVeiga:2008zz} extends this framework to $1+3$ dimensions, unveiling the Gell-Mann--Ne'eman eightfold way structure with baryon octets and decuplets. The masses, $M = -3 \ln \kappa - 3\kappa^3/4 + \kappa^6 r(\kappa)$, include an octet-decuplet splitting of $3\kappa^6/4 + \mathcal{O}(\kappa^7)$. The obtained spectral results for the one-baryon states exhibit quark confinement up to the corresponding energy threshold. As above, $r(\kappa)$ captures the contributions to the mass that arise from the $\kappa^6$ term in the series expansion, refining the mass prediction beyond the simple terms such as $-3 \ln \kappa$ and $-3 \kappa^3/4$. Here, the Hilbert space is enriched by including symmetries from both $SU(3)_c$ gauge and $SU(3)_f$ flavor groups, besides spin, and the isolated baryonic spectrum is rigorously connected to gauge-invariant states. Similar results have also been obtained for the mesonic sector of $\mathcal H$ \cite{FranciscoNeto:2008xbx, FranciscoNeto:2008ubr}. Two-particle baryonic bound states such as the deuterum were obtained e.g. in Refs. \cite{BS1,BS2}.
\end{remark}

In the Euclidean spacetime path integral formulation on the lattice, the expectation value of a given observable $\mathcal{O}$ is known to be \cite{Gattringer:2010zz,Montvay:1994cy}
\bea
\expval{\mathcal{O}}_{f,g}=\frac{1}{Z_{\Lambda}}\int d\psi d\bar{\psi}d\mu(g)\,\mathcal{O}\exp\left(-S_g-S_f\right),
\eea
where the partition function $Z_\Lambda$, to be thought of as a normalization, takes the form 
\bea
Z_\Lambda=\expval{1}_{f,g}=\int d\psi d\bar{\psi}d\mu(g) \exp\left(-S_g-S_f\right).
\eea
The integral measures are product measures, i.e.,
\bea \label{prod. measures}
d\psi d\bar{\psi}=\prod_{x,\,a,\,\alpha,\,f} d\psi_{a\alpha f}(x)d\bar{\psi}_{a\alpha f}(x),\quad d\mu(g)=\prod_{x,\,\mu} d\mu(g_{xx_\mu}).
\eea
where $d\mu(g)$ is a Haar measure.

Here, we are interested in a kind of nested integral in which we treat the fermionic and gauge parts separately (see \cite{Gattringer:2010zz}), though one still should be cautious regarding the additional contribution for the gauge part coming from the Fermi part. We will have more to say about this later. For the moment, we give the formula
\bea \label{Expval O_f,g}
\expval{\mathcal{O}}_{f,g}=\frac{1}{Z_{\Lambda}}\int d\mu(g)\left\langle \mathcal{O}\right\rangle _{f}Z_{f}f(U_{b})\exp\left(-S_{g}\right),
\eea
where $Z_f$ and $\expval{\mathcal{O}}_f$ denote the fermionic part of the partition function and the expectation value, respectively. To be specific, 
\bea
Z_f=\int d\psi d\bar{\psi}\exp(-S_f)
\label{Zfdef}
\eea
 and 
 \bea
 \expval{\mathcal{O}}_f=\frac{1}{Z_f}\,\int d\psi d\bar{\psi}\mathcal{O}\exp(-S_f)
 \label{Ofdef}
 \eea
 Moreover, the gauge part will be treated in a strong coupling regime that corresponds to $\exp(-S_g)\approx 1$. Now using (\ref{F-K}) and (\ref{Expval O_f,g}), what is to be computed for the expectation value $\expval{\mathcal{O}FG}$ should be clear. In what follows, we set \cite{FariadaVeiga:2004rf}
\bea
d\mu_C\left(\psi,\bar{\psi}\right)=\frac{1}{\mathcal{N}}d\psi d\bar{\psi}\,\exp\left(-a^{2}\sum_{x\in\Lambda}\bar{\psi}_{a\alpha f}(x)M_{\alpha\beta}\psi_{a\beta f}(y)\right),
\eea
is Gaussian with normalization ${\cal N}$ and covariance matrix $M$, given in (\ref{mass_term}), such that the Fermi integral between two independent variables $\psi$ and $\bar{\psi}$ takes the form
\bea \label{Fermi int.}
\int d\mu_C\left(\psi,\bar{\psi}\right)\,\psi_{a\alpha f}(x)\bar{\psi}_{b\beta h}(y)=\Xi\,\delta_{ab}\delta_{\alpha\beta}\delta_{fh}\delta(x-y),
\eea
with
\bea
\Xi\doteq a^{-2}\left(m+\frac{2\kappa}{a}\right)^{-1}.
\label{Xidef}
\eea
Here, we preserve the continuum Dirac delta notation for the discrete Kronecker delta in coordinate points.
%%%%%%%%%%%%%%%%%%%%%%%%%%%%%%%%%%%%%%%%%%%%%%%%%%%%%%%%%%%%%
%%%%%%%%%%%%%%%%%%%%%%%%%%%%%%%%%%%%%%%%%%%%%%%%%%%%%%%%%%%%%
\section{Quantum Yang-Mills charges on the lattice} \label{sect. main}
%%%%%%%%%%%%%%%%%%%%%%%%%%%%%%%%%%%%%%%%%%%%%%%%%%%%%%%%%%%%%
%%%%%%%%%%%%%%%%%%%%%%%%%%%%%%%%%%%%%%%%%%%%%%%%%%%%%%%%%%%%%
\subsection{Lattice charge operator}
%%%%%%%%%%%%%%%%%%%%%%%%%%%%%%%%%%%%%%%%%%%%%%%%%%%%%%%%%%%%%
%%%%%%%%%%%%%%%%%%%%%%%%%%%%%%%%%%%%%%%%%%%%%%%%%%%%%%%%%%%%%
Adapting the classical matter current, which is a Lie-algebra-valued function (as discussed in \cite{Ferreira:2012aj}), and incorporating the conjugation by Wilson lines, we express the temporal component of the current in (\ref{ymeqs}) as follows. For convenience, we will use $W(x)=W(x)^{-1}$ in (\ref{chargebeta1}) and (\ref{chargebeta2b}) 
\bea \label{current J_0}
J_0(x)=\sum_j J^j_0(x) W(x)\tau^j W(x)^{-1},
\eea
with the scalar coefficient $J^j_0(x)$, hereafter referred to simply as the coefficient, being
\bea \label{lat_current}
J_{0}^{j}(x)&\doteq &\frac{1}{2}\bar{\psi}_{c\xi f}(x_{0}^{+})\left(1+\gamma_{0}\right)_{\xi\eta}\left(g_{x_{0}^{+}x}\right)_{cd}\left(\tau^{j}\right)_{de}\psi_{e\eta f}\left(x\right)\nn\\
&&-\frac{1}{2}\bar{\psi}_{c\xi f}(x)\left(1-\gamma_{0}\right)_{\xi\eta}\left(\tau^{j}\right)_{cd}\left(g_{xx_{0}^{+}}\right)_{de}\psi_{e\eta f}\left(x_{0}^{+}\right),
\eea
where a summation is considered over repeated indices. The above coefficient exhibits the following characteristics
\begin{itemize}    
    \item It is covariantly conserved in the \emph{naive} continuum limit \cite{Gattringer:2010zz}. For a proof, see appendix \ref{appndx. cont. lim.}.
    \item It reduces to the usual $U(1)$ coefficient on the lattice (cf. \cite{Montvay:1994cy}), and the usual classical one\footnote{The quantum current is expected to involve some correction terms arising from renormalization and operator ordering effects, see e.g. \cite{Rejzner:2013ak}} in the continuum limit.
    \item While the coefficient transforms locally under the adjoint representation induced by the local gauge transformations, the current (\ref{current J_0}) transforms globally.
\end{itemize}

As such, the current given in (\ref{lat_current}) should not be interpreted as a physical quantum current.
Moreover, our formulation does not incorporate renormalization, as we do not attempt to define a fully consistent quantum current in the ultraviolet limit $a\searrow 0$ here.

In this context, we aim to explore the quantum corrections introduced by the conjugation with Wilson lines and the trace over color indices, a concept initially proposed by one of us \cite{Alvarez:1997ma,Alvarez:2009dt}. Wilson lines play a pivotal role in non-Abelian gauge theories, as they seem to encapsulate key quantum effects, such as confinement \cite{Wilson:1974sk}, and the behavior of gauge fields in a lattice framework \cite{Polyakov:1975rs}. Additionally, they provide a framework for understanding the gauge-invariant properties of electric and magnetic fluxes and their role in non-Abelian gauge theories \cite{tHooft:1979rtg}. The path-dependence of gauge transformations, which is crucial for describing these effects, is explored further in the context of Yang-Mills theory \cite{Dolan:1980qw}. By incorporating these elements, we hope to capture some of the nontrivial quantum features that arise in lattice gauge theories. For a more detailed discussion of this construction and its implications, we refer the reader to \cite{Alvarez:1997ma, Alvarez:2009dt, Ferreira:2011ed, Ferreira:2012aj}.

Using (\ref{current J_0}) and (\ref{lat_current}) for the direct implementation of the charge operator (\ref{chargebeta2b}) on the $G={\mathrm SU}(N)$ lattice model yields,
\bea \label{charge op. lattice}
Q^{(2)}_\ell(x,y)\doteq a^2\sum_{x,y\in \Lambda}\sum_{j,k}J^j_0(x)J^k_0(y)\,{\rm Tr}\Biggl( W(x)\tau^j W(x)^{-1}W(y)\tau^k W(y)^{-1}\Biggr),
\eea
where the index $\ell$ denotes the lattice version of $Q_M^{(2)}$, apart from the overall multiplicative constant $M(ig)^2/2$. We denote by Tr the trace operation over a representation of the gauge group $G$, which can be chosen arbitrarily, but preferably faithfully. In addition, the Wilson line is now understood as the (ordered) product of bond variables, specifically,
\bea
W(x)=g_{x_0x_1}g_{x_1x_2}\ldots g_{x_{n-1}x_n},
\eea
provided the path in (\ref{chargebeta2b}) splits into the \emph{spatial} intervals $x_0x_1$, $x_1x_2$, $\ldots$, $x_{n-1}x_n$ of the size of a lattice spacing $a$. The operator (\ref{charge op. lattice}) is the most important in our studies, since it leads to non zero contributions of the expectation values, hence potentially interesting for the charges, referred to as the eigenvalues of the operator (\ref{chargebeta2b}) in classical investigation. The procedure for implementing the higher-order operators of the series (\ref{chargeexpansion}) is analogous to that used in the construction of (\ref{charge op. lattice}).
%%%%%%%%%%%%%%%%%%%%%%%%%%%%%%%%%%%%%%%%%%%%%%%%%%%
%%%%%%%%%%%%%%%%%%%%%%%%%%%%%%%%%%%%%%%%%%%%%%%%%%%
\subsection{State vector analysis: expectation values for mesons and baryons}
%%%%%%%%%%%%%%%%%%%%%%%%%%%%%%%%%%%%%%%%%%%%%%%%%%%
%%%%%%%%%%%%%%%%%%%%%%%%%%%%%%%%%%%%%%%%%%%%%%%%%%%
Our goal is to investigate the quantum properties of the charge operator introduced above. Specifically, we will focus on the expectation values of the charge operator, evaluated with respect to the composite meson and baryon field vectors within the (quantum mechanical) Hilbert space $\mathcal{H}$.

\begin{remark}
We observe that meson fields are well defined for both gauge groups $G = \mathrm{SU}(2)$ and $G = \mathrm{SU}(3)$, corresponding to color indices taking values in $\{1, 2\}$ and $\{1, 2, 3\}$, respectively. We treat both cases in the discussion below. However, when referring to baryon fields, we restrict the gauge group to $G = \mathrm{SU}(3)$, as their construction involves the totally antisymmetric Levi-Civita symbol $\epsilon_{abc}$, which requires three distinct color indices (see~(\ref{vect. baryons})).
\end{remark}

The gauge invariant meson composite field vectors are given by \cite{Montvay:1994cy}
\bea \label{vect. mesons}
F=\bar{\psi}_{c\alpha_{1}f_{1}}(y)\psi_{c\alpha_{2}f_{2}}(y),\qquad G=\bar{\psi}_{d\beta_{1}h_{1}}\left(y_{\rho}^{\pi}\right)\psi_{d\beta_{2}h_{2}}\left(y_{\rho}^{\pi}\right).
\eea
\begin{remark} \label{rmrk.spin}
For computations, restrictions on the spin indices in (\ref{vect. mesons}) will be applied. Specifically, we will use $\alpha_1 \neq \alpha_2$ and $\beta_1 \neq \beta_2$, as mesons are physically composed of a quark and an antiquark paired appropriately.
\end{remark}
The gauge invariant baryon composite field vectors are given by \cite{Montvay:1994cy,FariadaVeiga:2004rf}
\bea \label{vect. baryons}
&F=\epsilon_{a_{1}a_{2}a_{3}}\psi_{a_{1}\alpha_{1}f_{1}}(y)\psi_{a_{2}\alpha_{2}f_{2}}(y)\psi_{a_{3}\alpha_{3}f_{3}}(y),\nn\\ &G=\epsilon_{b_{1}b_{2}b_{3}}\bar{\psi}_{b_{1}\beta_{1}h_{1}}\left(y_{\rho}^{\pi}\right)\bar{\psi}_{b_{2}\beta_{2}h_{2}}\left(y_{\rho}^{\pi}\right)\bar{\psi}_{b_{3}\beta_{3}h_{3}}\left(y_{\rho}^{\pi}\right).
\eea
%%%%%%%%%%%%%%%%%%%%%%%%%%%%%%%%%%%%%%%%%%%%%%%%%%%
%%%%%%%%%%%%%%%%%%%%%%%%%%%%%%%%%%%%%%%%%%%%%%%%%%%
\subsubsection{Matter part}
%%%%%%%%%%%%%%%%%%%%%%%%%%%%%%%%%%%%%%%%%%%%%%%%%%%
%%%%%%%%%%%%%%%%%%%%%%%%%%%%%%%%%%%%%%%%%%%%%%%%%%%

\subsubsection*{Mesons} Using Wick's theorem (see \cite{Gattringer:2010zz,Montvay:1994cy} for exposure on the lattice), in particular, (\ref{Fermi int.}) along with (\ref{prod. measures}), the second and the third term, which are cross terms of the product $J_{0}^{j}(x)J_{0}^{k}(x)$ with $\mathcal{O}(\kappa^0)$ yields the same results as follows:
\bea \label{expval_mesons}
\expval{J_{0}^{j}(x)J_{0}^{k}(x)FG}_f=\frac{1}{2}\frac{\Xi^{4}}{Z_{f}}\left(\tau^{j}\tau^{k}\right)_{d_{1}d_{1}}E_{f}^{\pm},
\eea
where $Z_f$ and $\Xi$ are defined respectively in (\ref{Zfdef}) and (\ref{Xidef}), and  $E_{f}^{\pm}$, for mesons, is given by 
\bea \label{E_f mesons}
E^\pm_f=\left(1\pm\gamma_{0}\right)_{\beta_{2}\alpha_{1}}\left(1\mp\gamma_{0}\right)_{\alpha_{2}\beta_{1}}\delta_{f_{1}h_{2}}\delta_{f_{2}h_{1}}.
\eea
In the computation, we also used
\begin{itemize}
    \item $y=x,\,\rho=0,\,\pi=+1$ for $E^+_f$;
    \item $y=x_{0}^{+},\,\rho=0,\,\pi=-1$ for $E^-_f$.
\end{itemize}

Naturally, the total contribution will be doubled for each of the choices of the parameters listed above, basically for $y=x$ and $y=x^+_0$. Note that if the corresponding vector states (\ref{vect. mesons}) are not gauge invariant, then there is \emph{no} nonzero contribution in the above computation, see the relevant discussion below (\ref{expval_m2}). For more details on nonzero contributions, we refer the reader to appendix \ref{appndx. nonzeros}.

\subsubsection*{Baryons}

As above, for the first and the fourth term (which is symmetric in spacetime points---except for the considered mesons) of the product $J^j_0(x)J^k_0(x)$ gives
\bea \label{expval_bar}
\expval{J_{0}^{j}(x)J_{0}^{k}(x)FG}_f=\frac{3a}{4}\frac{\Xi^{6}}{Z_{f}}\left(\tau^{j}\tau^{k}\right)_{d_{1}d_{1}} E_{f}^{\pm},
\eea
recalling that $Z_f$ and $\Xi$ are defined respectively in (\ref{Zfdef}) and (\ref{Xidef}). Here  $E_{f}^{\pm}$, for baryons, is given by
\bea \label{E_f baryons}
E_{f}^{\pm}&\doteq&
\resizebox{0.75\hsize}{!}{$\left[\left(1\pm\gamma_{0}\right)_{\alpha_{1}\beta_{1}}\left(1\pm\gamma_{0}\right)_{\alpha_{2}\beta_{2}}\delta_{f_{1}h_{1}}\delta_{f_{2}h_{2}}+\left(1\pm\gamma_{0}\right)_{\alpha_{1}\beta_{2}}\left(1\pm\gamma_{0}\right)_{\alpha_{2}\beta_{1}}\delta_{f_{1}h_{2}}\delta_{f_{2}h_{1}}\right]\left(1\pm\gamma_{0}\right)_{\alpha_{3}\beta_{3}}\delta_{f_{3}h_{3}}$}\nn\\
&&\resizebox{0.75\hsize}{!}{$+\left[\left(1\pm\gamma_{0}\right)_{\alpha_{1}\beta_{1}}\left(1\pm\gamma_{0}\right)_{\alpha_{2}\beta_{3}}\delta_{f_{1}h_{1}}\delta_{f_{2}h_{3}}+\left(1\pm\gamma_{0}\right)_{\alpha_{1}\beta_{3}}\left(1\pm\gamma_{0}\right)_{\alpha_{2}\beta_{1}}\delta_{f_{1}h_{3}}\delta_{f_{2}h_{1}}\right]\left(1\pm\gamma_{0}\right)_{\alpha_{3}\beta_{2}}\delta_{f_{3}h_{2}}$}\nn\\
&&\resizebox{0.75\hsize}{!}{$+\left[\left(1\pm\gamma_{0}\right)_{\alpha_{1}\beta_{2}}\left(1\pm\gamma_{0}\right)_{\alpha_{2}\beta_{3}}\delta_{f_{1}h_{2}}\delta_{f_{2}h_{3}}+\left(1\pm\gamma_{0}\right)_{\alpha_{1}\beta_{3}}\left(1\pm\gamma_{0}\right)_{\alpha_{2}\beta_{2}}\delta_{f_{1}h_{3}}\delta_{f_{2}h_{2}}\right]\left(1\pm\gamma_{0}\right)_{\alpha_{3}\beta_{1}}\delta_{f_{3}h_{1}}$}.
\eea
Here, the specific spacetime points and the associated parameters are chosen to be either of the following options, with $y=\tilde{x}^+_0$ for what we refer to as the first term (see (\ref{exp_bar})) and $y=\tilde{x}$ for what we refer to as the fourth term:
\begin{itemize}
    \item $y=\tilde{x}_{0}^{+}$, $\rho=0,\pi=-1$, $x=y,\mu=\rho$ , $\epsilon=\pi$  corresponds to $E^+_f$.
    \item $y=\tilde{x}$, $\rho=0,\pi=+1$, $x=y,\mu=\rho$ , $\epsilon=\pi$ corresponds to $E^-_f$.
\end{itemize}
\begin{remark} \label{commt. psi-psibar}
Here, nonzero contributions from two different terms of the product $J_{0}^{j}(x)J_{0}^{k}(x)$ for the same choice of $y$ are not possible, since baryons (\ref{vect. baryons}) are composed of $\psi$'s or $\bar{\psi}$'s, unlike mesons, which contain a mixture of both. 
\end{remark}
The argument concerning the non-Gauge invariant state vector is slightly subtle here and is treated separately at the end of appendix \ref{appndx. nonzeros}.
%%%%%%%%%%%%%%%%%%%%%%%%%%%%%%%%%%%%%%%%%%%%%%%%
%%%%%%%%%%%%%%%%%%%%%%%%%%%%%%%%%%%%%%%%%%%%%%%%
\subsubsection{Gauge part}
%%%%%%%%%%%%%%%%%%%%%%%%%%%%%%%%%%%%%%%%%%%%%%%%
%%%%%%%%%%%%%%%%%%%%%%%%%%%%%%%%%%%%%%%%%%%%%%%%
Here, both the meson and baryon composite fields yield the same outcome. Since the covariance matrix is diagonal in spacetime points, cf. Isserlis-Wick theorem, the gauge integral is expected to produce trivial values. Specifically, when taking the trace over the color indices for the charge operator, the diagonal terms in spacetime points simplify directly to an expression that is free from the conjugation by Wilson lines.

For the bond variables appearing in the product $J_{0}^{j}(x)J_{0}^{k}(x)$, cf. (\ref{lat_current}), the gauge integral becomes trivial after performing the Fermi integral in the case of mesons. However, for baryons, the manipulation is slightly more involved. Specifically, the result simplifies in a manner analogous to mesons but requires additional care in handling the gauge integral by leveraging the following determinant expression for the bond variables (see appendix \ref{appndx. nonzeros}):
\bea \label{det g}
\epsilon_{c_{1}c_{2}a}\left(g_{\tilde{x}_{0}^{+}\tilde{x}}\right)_{c_{1}d_{1}}\left(g_{\tilde{x}_{0}^{+}\tilde{x}}\right)_{c_{2}d_{2}}\left(g_{\tilde{x}_{0}^{+}\tilde{x}}\right)_{ab}=\epsilon_{d_{1}d_{2}b}\det g,
\eea
such that
\bea \label{epsilons}
\epsilon_{e_{1}e_{2}b}\epsilon_{d_{1}d_{2}b}=\delta_{e_{1}d_{1}}\delta_{e_{2}d_{2}}-\delta_{e_{1}d_{2}}\delta_{e_{2}d_{1}},
\eea
gives the desired result (\ref{expval_bar}). The first term of the product $J^j_0(x)J^k_0(x)$, for example, involves the term
\bea\label{tau^2}
\epsilon_{c_{1}c_{2}a}\epsilon_{e_{1}e_{2}b}\left(g_{\tilde{x}_{0}^{+}\tilde{x}}\right)_{c_{1}d_{1}}\left(g_{\tilde{x}_{0}^{+}\tilde{x}}\right)_{c_{2}d_{2}}\left(g_{\tilde{x}_{0}^{+}\tilde{x}}\right)_{ab}\left(\tau^{j}\right)_{d_{1}e_{1}}\left(\tau^{k}\right)_{d_{2}e_{2}},
\eea
which upon using (\ref{det g}) and (\ref{epsilons}) immediately gives $\left(\tau^{j}\tau^{k}\right)_{d_{1}d_{1}}$.

On the other hand, if one decides to carry out the gauge integral first, all that is required is the following fundamental integration orthogonality formula, arising from Peter-Weyl theorem \cite{Creutz:1978ub,Gattringer:2010zz, simon1996representations},
\bea
\int d\mu(g)\,g_{i_{1}j_{1}}g_{k_{1}l_{1}}^{-1}=\frac{1}{N}\delta_{i_{1}l_{1}}\delta_{j_{1}k_{1}},
\eea
where $N=2,3$ refers to the gauge group $SU(N)$.
%%%%%%%%%%%%%%%%%%%%%%%%%%%%%%%%%%%%%%%%%%%%%%%%
%%%%%%%%%%%%%%%%%%%%%%%%%%%%%%%%%%%%%%%%%%%%%%%%
\subsubsection{Total contribution: summing expectation values}
%%%%%%%%%%%%%%%%%%%%%%%%%%%%%%%%%%%%%%%%%%%%%%%%
%%%%%%%%%%%%%%%%%%%%%%%%%%%%%%%%%%%%%%%%%%%%%%%%
The total charge, obtained from the expectation values of the charge operator $Q^{(2)}_\ell(x)$ (\ref{charge op. lattice}), is expressed as a series comprising $n$ identical terms of the type given in (\ref{expval_mesons}) for mesons and (\ref{expval_bar}) for baryons, along with additional contributions from the gauge sector. Specifically, the total (quantum) charge—i.e., the sum of expectation values for \emph{mesons} or \emph{baryons}—takes the form, cf. (\ref{Expval O_f,g})
\bea
\expval{Q^{(2)}_\ell FG}_{f,g}=\frac{a^2}{Z_\Lambda}\sum_{j,k} n \Tr(\tau^j \tau^k) \expval{J_{0}^{j}(x)J_{0}^{k}(x)FG}_f Z_f,
\label{finalchargevalue}
\eea
where the sum extends over $n$ lattice points $x = y = x_1, \ldots, x_n \in \Lambda$. The composite fields $F$ and $G$, as well as the fermionic contribution $\expval{\cdot}_f$ to the expectation value, depend on whether one is considering mesons (\ref{vect. mesons}) and (\ref{expval_mesons}) or baryons (\ref{vect. baryons}) and (\ref{expval_bar}).

In the limit \( n \to \infty \)—the infinite volume or thermodynamic limit \cite{Gattringer:2010zz}—the series in question along with the partition function \( Z_\Lambda \) in the denominator diverges. Since we prefer to work on a finite nonzero lattice spacing, we do not employ the standard renormalization techniques commonly used in lattice QCD to handle such divergences. However, alternative approaches exist. In particular, the multiscale polymer expansion \cite{Simon1993, Glimm:1987ng} offers a systematic framework to address these issues, ensuring that divergences can be removed safely, yielding finite, well-defined results. Other regularization techniques from quantum field theory can be used, and may be relevant for future investigations depending on the underlying model. 

The average appearing in (\ref{finalchargevalue}) are computed and given in Appendix \ref{sec:appendixfinalresults} for hadrons, in the case of lattice QCD models with two and three flavors. Note that our explicit computations are in agreement with the flavor symmetry of the model. By considering the hadronic particles wave functions in terms of quarks and anti-quarks, and their linear combinations, the values of the quantum charges of the hadrons follow. For instance, in the three flavor case and considering the pre-factors given in (\ref{expval_mesons}), the charge values for $\pi^{+}$ and $\pi^{-}$ coincide, and it vanishes for $\pi^0$. Regarding baryons and using (\ref{expval_bar}), we obtain that the charge value for $\Delta^{++}$ differs from $p$ and $n$ (proton and neutron), which are equal. 

The reason for the vanishing charge of $\pi^0$ is algebraic, and is due to the fact that its wave function is the linear combination $\pi^0\equiv \left(u\,{\bar u}- d\,{\bar d}\right)/\sqrt{2}$, whilst for $\pi^{\pm}$ the linear combination with a minus sign is not there, as $\pi^+\equiv u\,{\bar d}$, and $\pi^-\equiv d\,{\bar u}$. Indeed, using (\ref{expval_mesons}) and the tables in Appendix \ref{sec:appendixfinalresults} gives non-zero charge for $\pi^{\pm}$ and a vanishing charge for $\pi^0$.

We also considered gauge variant states for which we obtain vanishing charges.

\section{Conclusion} \label{sect. conclusion}
We constructed a two-dimensional Euclidean quantum gauge invariant Wilson lattice version of the gauge invariant charge using Wilson loops and developed in \cite{Ferreira:2011ed,Ferreira:2012aj,twoclassical}. The charge is given as an infinite series.
In order to understand the physical property measured by this charge and improve our knowledge about the role of local gauge invariance in quantum field models, we computed the quantum charge effects up to second order acting on non-gauge invariant states and hadrons in lattice QCD with two and three quark flavors. In the strong-coupling regime, we obtain a zero eigenvalue for the quantum charge for non-gauge invariant states and non-zero values for baryons (see (\ref{expval_bar})) and mesons (see (\ref{expval_mesons})). In a more realistic approach, we need to incorporate higher-order contributions to the charge. In addition, renormalization effects must also be taken into account to define the lattice QCD model. Whether or not our quantum charge is a measure of confinement is a question that deserves further analysis.

Of course it would be very desirable to understand the role of the charge $Q_M^{(2)}$ in a more phenomenological way, e.g. analyzing some standard decay processes as $\pi^0\rightarrow 2\,\gamma$,  $\pi^{+}\rightarrow \mu^{+}+\nu_{\mu}$, and $\Delta^{++}\rightarrow p+\pi^+$. There is a wide open pathway to explore on those lines. 

In addition, it would also be interesting to extend our analysis to the quantum version of the higher order charges $Q_M^{(j)}$, given in (\ref{chargeexpansion}). And, more importantly, to extend the analysis to the case of real QCD in $(3+1)$-dimensions, and verify if the real hadrons carry the quantum version of the charges constructed  in \cite{ferreira2025}. 

\section*{Acknowledgment}

LAF acknowledges the financial support of Fapesp
(Funda\c c\~ao de Amparo \`a Pesquisa do Estado de S\~ao Paulo) grant 2022/00808-7, and CNPq
(Conselho Nacional de Desenvolvimento Cient\'ifico e Tecnol\'ogico) grant 307833/2022-4. HM acknowledges the financial support of Fapesp grant 2021/10141-7. RM's work was partially supported by the FAPESP TT5 grant 2023/18074-2, and subsequently by the CNPq PDJ grant 168581/2023-0.

\appendix

\section{Non-zero contributions} \label{appndx. nonzeros}

For the reader's convenience, we provide the necessary details on the nonzero contributions arising from the fermionic components of the expectation values for mesons and baryons. To illustrate, we present a relevant tree-level diagram for mesons, highlighting an interesting case. For baryons, the tree-level diagrams follow a similar logic but are too cumbersome to present explicitly in the text. Instead, we outline the principal nonzero terms for baryons and provide a comprehensive list showing how these terms relate to other nonzero contributions. We will neglect the global normalization corresponding to the partition function $Z_\Lambda$ (see (\ref{Expval O_f,g})), which is present in all the analyzed cases. Hence, we will concentrate on the unnormalized expectation values.

\subsection*{Mesons}

Here we exemplify the explicit computation, since the tree diagram is smaller compared to that of baryons, and easy to demonstrate the main features.

We are interested in computing the Fermionic part of the second term, and the first cross term, again it is the term containing asymmetric $\psi$'s and $\bar{\psi}$'s with respect to spacetime points, of the product $J_{0}^{j}(x)J_{0}^{k}(x)$ at the order $\mathcal{O}(\kappa^0)$, in particular, regarding Fig. \ref{fig:tree-diag} below, we first examine the spacetime points and identify the potentially nonzero terms. There are a total of six terms, of which the following do not survive:

\begin{itemize}
    \item (4)(1)(1): This term involves contractions with spacetime points $y$ and $y^\pi_\rho$, which are distinct by construction, specifically $\delta(y - y^\pi_\rho) = 0$.
    \item (4)(2)(1): This term does not survive for the same reason as the previous one.
    \item (4)(2)(2): This term involves contractions with $x$ and $x^+_0$, which results in a product that is zero.
    \item (4)(3)(2): This term does not survive because, for it to be nonzero, one would have to set $y = x$, which subsequently affects other contractions in the subbranch, yielding trivial zeros, $\delta(x - x^+_0)$ and $\delta(y^\pi_\rho-y)$.
\end{itemize}

After analyzing all the possible combinations, that accounts to analyzing three more tree diagrams, we arrive at the following potentially nonzero terms:

\begin{enumerate}
    \item  For $y=x$, $\rho=0$ and $\pi=+1$
\[
(2)(1)(1):\delta_{c_{1}e_{2}}\delta_{\xi_{1}\eta_{2}}\delta_{fh}\delta_{c_{2}e_{1}}\delta_{\xi_{2}\eta_{1}}\delta_{hf}\delta_{cc}\delta_{\alpha_{1}\alpha_{2}}\delta_{f_{1}f_{2}}\delta_{dd}\delta_{\beta_{1}\beta_{2}}\delta_{h_{1}h_{2}},
\]
\[
(2)(2)(1):\delta_{c_{1}e_{2}}\delta_{\xi_{1}\eta_{2}}\delta_{fh}\delta_{c_{2}c}\delta_{\xi_{2}\alpha_{2}}\delta_{hf_{2}}\delta_{ce_{1}}\delta_{\alpha_{1}\eta_{1}}\delta_{f_{1}f}\delta_{dd}\delta_{\beta_{1}\beta_{2}}\delta_{h_{1}h_{2}},
\]
\[
(4)(1)(2):\delta_{c_{1}d}\delta_{\xi_{1}\beta_{2}}\delta_{fh_{2}}\delta_{c_{2}e_{1}}\delta_{\xi_{2}\eta_{1}}\delta_{hf}\delta_{cc}\delta_{\alpha_{1}\alpha_{2}}\delta_{f_{1}f_{2}}\delta_{de_{2}}\delta_{\beta_{1}\eta_{2}}\delta_{h_{1}h},
\]
\[
(4)(3)(1):\delta_{c_{1}d}\delta_{\xi_{1}\beta_{2}}\delta_{fh_{2}}\delta_{c_{2}c}\delta_{\xi_{2}\alpha_{2}}\delta_{hf_{2}}\delta_{ce_{1}}\delta_{\alpha_{1}\eta_{1}}\delta_{f_{1}f}\delta_{de_{2}}\delta_{\beta_{1}\eta_{2}}\delta_{h_{1}h}.
\]

\item For $y=x^+_0$, $\rho=0$ and $\pi=-1$
\[
(3)(1)(1):\delta_{c_{1}c}\delta_{\xi_{1}\alpha_{2}}\delta_{ff_{2}}\delta_{c_{2}e_{1}}\delta_{\xi_{2}\eta_{1}}\delta_{hf}\delta_{ce_{2}}\delta_{\alpha_{1}\eta_{2}}\delta_{f_{1}h}\delta_{dd}\delta_{\beta_{1}\beta_{2}}\delta_{h_{1}h_{2}}
\]
\[
(3)(3)(2):\delta_{c_{1}c}\delta_{\xi_{1}\alpha_{2}}\delta_{ff_{2}}\delta_{c_{2}d}\delta_{\xi_{2}\beta_{2}}\delta_{hh_{2}}\delta_{ce_{2}}\delta_{\alpha_{1}\eta_{2}}\delta_{f_{1}h}\delta_{de_{1}}\delta_{\beta_{1}\eta_{1}}\delta_{h_{1}f}
\]

\end{enumerate}

However, with regards to remark \ref{rmrk.spin} we see that $\delta_{\alpha_{1}\alpha_{2}}=\delta_{\beta_{1}\beta_{2}}=0$. Thus, the only surviving terms are (4)(3)(1) and (3)(3)(2). While the former produces
\bea \label{expval_m1}
\expval{J_{0}^{j}(x)J_{0}^{k}(x)FG}_{f}&=&\frac{1}{4Z_{f}}\int d\mu_{C}(\psi,\bar{\psi})\Biggl\{ \psi_{d\beta_{2}h_{2}}\left(y_{\rho}^{\pi}\right)\bar{\psi}_{c_{1}\xi_{1}f}\left(x_{0}^{+}\right)\psi_{c\alpha_{2}f_{2}}(y)\bar{\psi}_{c_{2}\xi_{2}h}(x)\nn\\
&&\psi_{e_{1}\eta_{1}f}\left(x\right)\bar{\psi}_{c\alpha_{1}f_{1}}(y)\psi_{e_{2}\eta_{2}h}\left(x_{0}^{+}\right)\bar{\psi}_{d\beta_{1}h_{1}}\left(y_{\rho}^{\pi}\right)\Biggr\}\nn\\
&&\left(1+\gamma_{0}\right)_{\xi_{1}\eta_{1}}\left(g_{x_{0}^{+}x}\right)_{c_{1}d_{1}}\left(\tau^{j}\right)_{d_{1}e_{1}}\left(1-\gamma_{0}\right)_{\xi_{2}\eta_{2}}\left(\tau^{k}\right)_{c_{2}d_{2}}\left(g_{xx_{0}^{+}}\right)_{d_{2}e_{2}}\nn\\
&=&\frac{\Xi^{4}}{4Z_{f}}\left(1+\gamma_{0}\right)_{\beta_{2}\alpha_{1}}\left(1-\gamma_{0}\right)_{\alpha_{2}\beta_{1}}\delta_{f_{1}h_{2}}\delta_{f_{2}h_{1}}\left(\tau^{j}\tau^{k}\right)_{d_{1}d_{1}},
\eea
the latter gives
\bea \label{expval_m2}
\expval{J_{0}^{j}(x)J_{0}^{k}(x)FG}_{f}&=&\frac{1}{4Z_{f}}\int d\mu_{C}(\psi,\bar{\psi})\Biggl\{ \psi_{c\alpha_{2}f_{2}}\left(y\right)\bar{\psi}_{c_{1}\xi_{1}f}\left(x_{0}^{+}\right)\psi_{d\beta_{2}h_{2}}\left(y_{\rho}^{\pi}\right)\bar{\psi}_{c_{2}\xi_{2}h}(x)\nn\\
&&\psi_{e_{2}\eta_{2}h}\left(x_{0}^{+}\right)\bar{\psi}_{c\alpha_{1}f_{1}}\left(y\right)\psi_{e_{1}\eta_{1}f}\left(x\right)\bar{\psi}_{d\beta_{1}h_{1}}\left(y_{\rho}^{\pi}\right)\Biggr\}\nn\\
&&\left(1+\gamma_{0}\right)_{\xi_{1}\eta_{1}}\left(g_{x_{0}^{+}x}\right)_{c_{1}d_{1}}\left(\tau^{j}\right)_{d_{1}e_{1}}\left(1-\gamma_{0}\right)_{\xi_{2}\eta_{2}}\left(\tau^{k}\right)_{c_{2}d_{2}}\left(g_{xx_{0}^{+}}\right)_{d_{2}e_{2}}\nn\\
&=&\frac{\Xi^{4}}{4Z_{f}}\left(1+\gamma_{0}\right)_{\alpha_{2}\beta_{1}}\left(1-\gamma_{0}\right)_{\beta_{2}\alpha_{1}}\delta_{f_{1}h_{2}}\delta_{f_{2}h_{1}}\left(\tau^{j}\tau^{k}\right)_{d_{1}d_{1}}.
\eea
The second cross term gives the same results as in (\ref{expval_m1}) and (\ref{expval_m2}) but with $y$'s and the corresponding parameters interchanged. As such the overall contribution yielding (\ref{expval_mesons}) is twice of the values given in (\ref{expval_m1}) and (\ref{expval_m2}).

Now, suppose the meson composite field (\ref{vect. mesons}) is not gauge-invariant, meaning we take $c \neq c'$. Following the above computation, we eventually obtain $(\tau^k\tau^j)_{cc'}$. However, the normalization of the underlying Lie algebra elements involves (up to a constant) $\delta_{cc'}$, which requires $c = c'$ for a nonzero contribution. Doing it the other way around—i.e., taking either $d \neq d'$ or restriction on both indices ($c$ and $d$) simultaneously—ultimately introduces an additional factor $\delta_{dd'}$, which contributes to a nonzero result only if $d = d'$. In other words, the expectation value for non-gauge-invariant $\bar{\psi}\psi$ vectors vanishes.

The absence of contributions from non-gauge-invariant field vectors is also evident when applying the hyperplane decoupling expansion to identify the physical particles, namely hadrons and glueballs \cite{FariadaVeiga:2008zz, Schor:1983aq}.

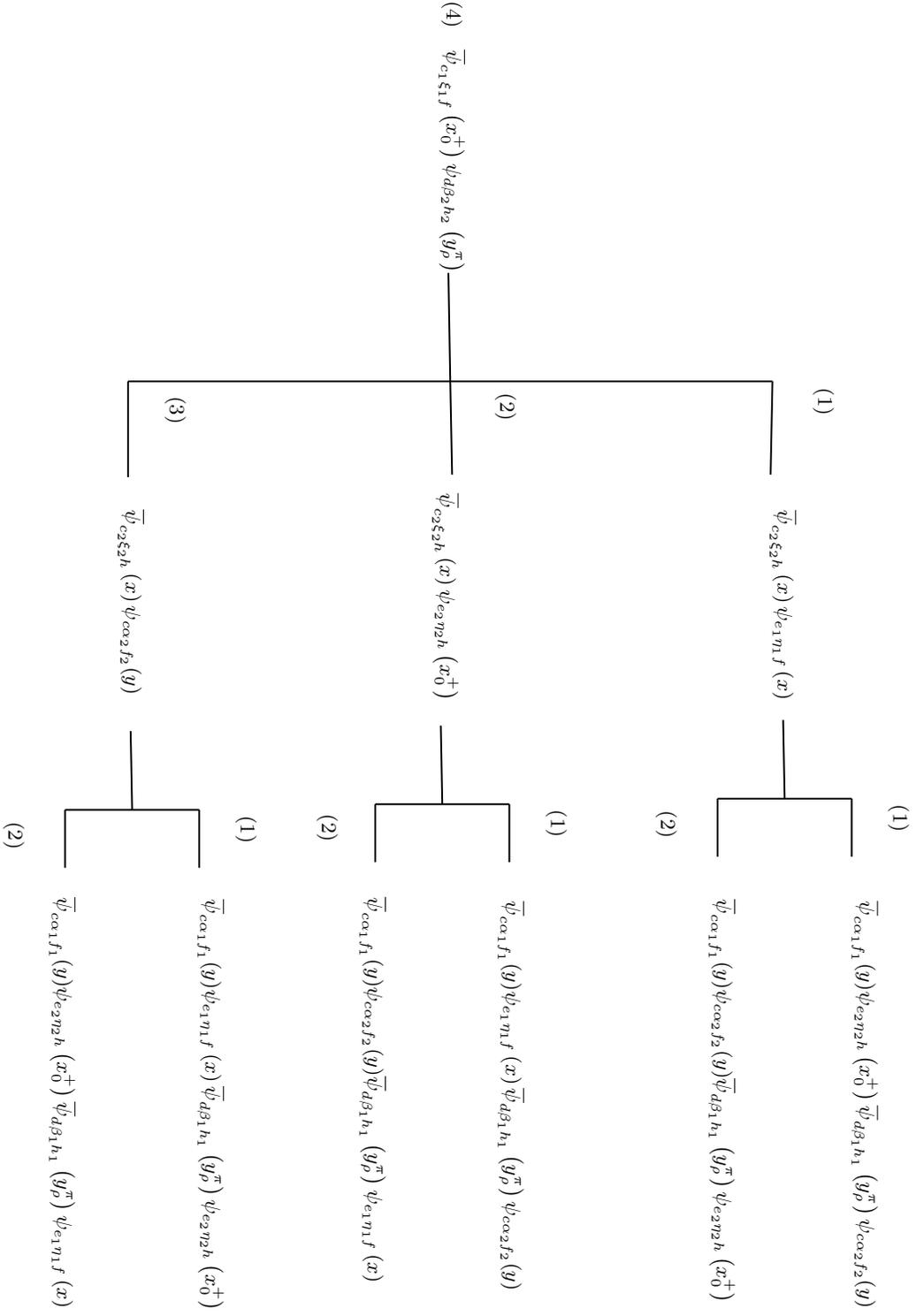
\begin{figure}\label{fig:tree-diag}
    \centering

    \begin{tikzpicture}[x=1pt,y=1pt,yscale=-0.8,xscale=1]

%Straight Lines [id:da371519547035831] 
\draw    (460.1,231.59) -- (184.41,231.59) ;
%Straight Lines [id:da9027611933637789] 
\draw    (322.26,231.59) -- (323.05,281.54) ;
%Straight Lines [id:da3828394236063821] 
\draw    (184.41,231.59) -- (184.41,282.88) ;
%Straight Lines [id:da32113836294733034] 
\draw    (321.46,173.54) -- (322.26,231.59) ;
%Straight Lines [id:da24967408918996647] 
\draw    (460.1,231.59) -- (459.3,281.54) ;
%Straight Lines [id:da03787093070677083] 
\draw    (464.67,412.88) -- (465.31,455) ;
%Straight Lines [id:da607194027474681] 
\draw    (493.99,455) -- (436.62,455) ;
%Straight Lines [id:da9417498576625865] 
\draw    (493.99,455) -- (493.99,486.05) ;
%Straight Lines [id:da47706449789852945] 
\draw    (436.62,455) -- (436.62,486.05) ;
%Straight Lines [id:da37471113580106974] 
\draw    (318.22,415.87) -- (318.85,457.99) ;
%Straight Lines [id:da7250276339883346] 
\draw    (347.54,457.99) -- (290.17,457.99) ;
%Straight Lines [id:da48835737791335054] 
\draw    (347.54,457.99) -- (347.54,489.03) ;
%Straight Lines [id:da6243365236067782] 
\draw    (290.17,457.99) -- (290.17,489.03) ;
%Straight Lines [id:da07540034577942234] 
\draw    (185.52,418.86) -- (186.16,460.97) ;
%Straight Lines [id:da6686908191604695] 
\draw    (214.84,460.97) -- (157.48,460.97) ;
%Straight Lines [id:da5546149265016982] 
\draw    (214.84,460.97) -- (214.84,492.02) ;
%Straight Lines [id:da5703451283449965] 
\draw    (157.48,460.97) -- (157.48,492.02) ;

% Text Node
\draw (506.75,508.36) node [anchor=north west][inner sep=0.75pt]  [font=\footnotesize,rotate=-90]  {$\overline{\psi }_{c\alpha _{1} f_{1}}( y) \psi _{e_{2} \eta _{2} h}\left(x_{0}^{+}\right)\overline{\psi }_{d\beta _{1} h_{1}}\left( y_{\rho }^{\pi }\right) \psi _{c\alpha _{2} f_{2}}( y)$};
% Text Node
\draw (445.61,507.41) node [anchor=north west][inner sep=0.75pt]  [font=\footnotesize,rotate=-90]  {$\overline{\psi }_{c\alpha _{1} f_{1}}( y) \psi _{c\alpha _{2} f_{2}}( y)\overline{\psi }_{d\beta _{1} h_{1}}\left( y_{\rho }^{\pi }\right) \psi _{e_{2} \eta _{2} h}\left(x_{0}^{+}\right)$};
% Text Node
\draw (355.36,508.45) node [anchor=north west][inner sep=0.75pt]  [font=\footnotesize,rotate=-90]  {$\overline{\psi }_{c\alpha _{1} f_{1}}( y) \psi _{e_{1} \eta _{1} f}\left(x\right)\overline{\psi }_{d\beta _{1} h_{1}}\left( y_{\rho }^{\pi }\right) \psi _{c\alpha _{2} f_{2}}( y)$};
% Text Node
\draw (296.22,506.23) node [anchor=north west][inner sep=0.75pt]  [font=\footnotesize,rotate=-90]  {$\overline{\psi }_{c\alpha _{1} f_{1}}( y) \psi _{c\alpha _{2} f_{2}}( y)\overline{\psi }_{d\beta _{1} h_{1}}\left( y_{\rho }^{\pi }\right) \psi _{e_{1} \eta _{1} f}\left(x\right)$};
% Text Node
\draw (226.9,507.41) node [anchor=north west][inner sep=0.75pt]  [font=\footnotesize,rotate=-90]  {$\overline{\psi }_{c\alpha _{1} f_{1}}( y) \psi _{e_{1} \eta _{1} f}\left(x\right)\overline{\psi }_{d\beta _{1} h_{1}}\left( y_{\rho }^{\pi }\right) \psi _{e_{2} \eta _{2} h}\left(x_{0}^{+}\right)$};
% Text Node
\draw (163.4,506.36) node [anchor=north west][inner sep=0.75pt]  [font=\footnotesize,rotate=-90]  {$\overline{\psi }_{c\alpha _{1} f_{1}}( y) \psi _{e_{2} \eta _{2} h}\left(x_{0}^{+}\right)\overline{\psi }_{d\beta _{1} h_{1}}\left( y_{\rho }^{\pi }\right) \psi _{e_{1} \eta _{1} f}\left(x\right)$};
% Text Node
\draw (330.04,26.21) node [anchor=north west][inner sep=0.75pt]  [font=\footnotesize,rotate=-90]  {$( 4)\quad\overline{\psi }_{c_{1} \xi _{1} f}\left(x_{0}^{+}\right) \psi _{d\beta _{2} h_{2}}\left( y_{\rho }^{\pi }\right)$};
% Text Node
\draw (487.54,233.65) node [anchor=north west][inner sep=0.75pt]  [font=\footnotesize,rotate=-90]  {$( 1)$};
% Text Node
\draw (351.59,236.35) node [anchor=north west][inner sep=0.75pt]  [font=\footnotesize,rotate=-90]  {$( 2)$};
% Text Node
\draw (210.26,238.05) node [anchor=north west][inner sep=0.75pt]  [font=\footnotesize,rotate=-90]  {$( 3)$};
% Text Node
\draw (471.72,299.06) node [anchor=north west][inner sep=0.75pt]  [font=\footnotesize,rotate=-90]  {$\overline{\psi }_{c_{2} \xi _{2} h}\left(x\right) \psi _{e_{1} \eta _{1} f}\left(x\right)$};
% Text Node
\draw (327.28,290.04) node [anchor=north west][inner sep=0.75pt]  [font=\footnotesize,rotate=-90]  {$\overline{\psi }_{c_{2} \xi _{2} h}\left(x\right) \psi _{e_{2} \eta _{2} h}\left(x_{0}^{+}\right)$};
% Text Node
\draw (192.72,298.74) node [anchor=north west][inner sep=0.75pt]  [font=\footnotesize,rotate=-90]  {$\overline{\psi }_{c_{2} \xi _{2} h}\left(x\right) \psi _{c\alpha _{2} f_{2}}( y)$};
% Text Node
\draw (519.67,456.52) node [anchor=north west][inner sep=0.75pt]  [font=\footnotesize,rotate=-90]  {$( 1)$};
% Text Node
\draw (420.37,459.86) node [anchor=north west][inner sep=0.75pt]  [font=\footnotesize,rotate=-90]  {$( 2)$};
% Text Node
\draw (373.22,459.51) node [anchor=north west][inner sep=0.75pt]  [font=\footnotesize,rotate=-90]  {$( 1)$};
% Text Node
\draw (275.34,461.85) node [anchor=north west][inner sep=0.75pt]  [font=\footnotesize,rotate=-90]  {$( 2)$};
% Text Node
\draw (240.52,462.5) node [anchor=north west][inner sep=0.75pt]  [font=\footnotesize,rotate=-90]  {$( 1)$};
% Text Node
\draw (141.22,465.84) node [anchor=north west][inner sep=0.75pt]  [font=\footnotesize,rotate=-90]  {$( 2)$};

\end{tikzpicture}
    
    \caption{A tree diagram for the fermionic part of the expectation values taken with respect to meson composite vector fields  with $y=x$, $\rho=0$ and $\pi=+1$.}
    
\end{figure}

\subsection*{Baryons}

Recalling that, here, the gauge group is ${\mathrm SU}(3)$ the product $J_{0}^{j}(x) J_{0}^{k}(x)$ at $\mathcal{O}(\kappa^1)$ yields (to avoid confusion with the spacetime points and flavor indices associated with the fields $\psi(x)$ and $\bar{\psi}(x)$ in the action (\ref{act. f}), and the fields appearing in the product $J_0^j(x) J_0^k(x)$, we will relabel the spacetime coordinates of the latter as $\tilde{x}$ and use capital letters, such as $F$, to denote the flavor indices of the former):

\bea \label{exp_bar}
\expval{J_{0}^{j}(\tilde{x})J_{0}^{k}(\tilde{x})FG}_f&=&-\frac{a}{8Z_{f}}\epsilon_{a_{1}a_{2}a_{3}}\epsilon_{b_{1}b_{2}b_{3}}\sum_{x,\mu,\epsilon}\int d\mu_{C}(\psi,\bar{\psi})\Bigl\{\bar{\psi}_{c_{1}\xi_{1}f}(\tilde{x}_{0}^{+})\psi_{e_{1}\eta_{1}f}\left(\tilde{x}\right)\bar{\psi}_{c_{2}\xi_{2}h}(\tilde{x}_{0}^{+})\nn\\
&&\psi_{e_{2}\eta_{2}h}\left(\tilde{x}\right)\bar{\psi}_{b_{1}\beta_{1}h_{1}}\left(y_{\rho}^{\pi}\right)\psi_{a_{1}\alpha_{1}f_{1}}(y)
\bar{\psi}_{b_{2}\beta_{2}h_{2}}\left(y_{\rho}^{\pi}\right)\psi_{a_{2}\alpha_{2}f_{2}}(y)\bar{\psi}_{b_{3}\beta_{3}h_{3}}\left(y_{\rho}^{\pi}\right)\nn\\
&&\psi_{a_{3}\alpha_{3}f_{3}}(y)\bar{\psi}_{a\alpha F}(x)\psi_{b\beta F}\left(x_{\mu}^{\epsilon}\right)\Bigr\}\left(1+\gamma_{0}\right)_{\xi_{1}\eta_{1}}\left(g_{\tilde{x}_{0}^{+}\tilde{x}}\right)_{c_{1}d_{1}}\left(\tau^{j}\right)_{d_{1}e_{1}}\nn\\
&&\left(1+\gamma_{0}\right)_{\xi_{2}\eta_{2}}\left(g_{\tilde{x}_{0}^{+}\tilde{x}}\right)_{c_{2}d_{2}}\left(\tau^{k}\right)_{d_{2}e_{2}}\Gamma_{\alpha\beta}^{\epsilon e_{\mu}}\left(g_{xx_{\mu}^{\epsilon}}\right)_{ab}	
\eea

Below, we list all possible nonzero terms along with their corresponding references for tracking within the tree diagram, as well as the chosen spacetime points and parameters used in the computation. The assignment of a reference number depends on the chosen combinatorial pattern, but once established, each term can be uniquely referenced within the diagram.\\

$\left(3\right)\left(3\right)\left(1\right)\left(1\right)\left(2\right), y=\tilde{x}_{0}^{+}, \rho=0,\pi=-1, x=y,\mu=\rho , \epsilon=\pi 
$
\bea \label{nonzero_b_1}
&&-\int d\mu_{C}(\psi,\bar{\psi})\Bigl\{ \psi_{a_{1}\alpha_{1}f_{1}}(y)\bar{\psi}_{c_{1}\xi_{1}f}(\tilde{x}_{0}^{+})\psi_{a_{2}\alpha_{2}f_{2}}(y)\bar{\psi}_{c_{2}\xi_{2}h}(\tilde{x}_{0}^{+})\psi_{e_{1}\eta_{1}f}\left(\tilde{x}\right)\bar{\psi}_{b_{1}\beta_{1}h_{1}}\left(y_{\rho}^{\pi}\right)\nn\\
&&\psi_{e_{2}\eta_{2}h}\left(\tilde{x}\right)\bar{\psi}_{b_{2}\beta_{2}h_{2}}\left(y_{\rho}^{\pi}\right)\psi_{b\beta F}\left(x_{\mu}^{\epsilon}\right)\bar{\psi}_{b_{3}\beta_{3}h_{3}}\left(y_{\rho}^{\pi}\right)\psi_{a_{3}\alpha_{3}f_{3}}(y)\bar{\psi}_{a\alpha F}(x)\Bigr\},
\eea
where we have included the overall sign arising from the permutations. Now, solving the integral using (\ref{Fermi int.}) along with (\ref{prod. measures}), then contracting the relevant flavour indices we arrive at
\bea \label{F-contractions b1}
-\Xi^{6}\left(\delta_{a_{1}c_{1}}\delta_{\alpha_{1}\xi_{1}}\right)\left(\delta_{a_{2}c_{2}}\delta_{\alpha_{2}\xi_{2}}\right)\left(\delta_{e_{1}b_{1}}\delta_{\eta_{1}\beta_{1}}\right)\left(\delta_{e_{2}b_{2}}\delta_{\eta_{2}\beta_{2}}\right)\left(\delta_{bb_{3}}\delta_{\beta\beta_{3}}\right)\left(\delta_{a_{3}a}\delta_{\alpha_{3}\alpha}\right)\delta_{f_{1}h_{1}}\delta_{f_{2}h_{2}}\delta_{f_{3}h_{3}},
\eea
thereby substituting the result in (\ref{exp_bar}) one immediately obtains:
\bea \label{B-33112}
&&-\frac{a}{8}\frac{\Xi^{6}}{Z_{f}}\epsilon_{c_{1}c_{2}a}\epsilon_{e_{1}e_{2}b}\left(1+\gamma_{0}\right)_{\alpha_{1}\beta_{1}}\left(1+\gamma_{0}\right)_{\alpha_{2}\beta_{2}}(1+\gamma_{0})_{\alpha_{3}\beta_{3}}\left(g_{\tilde{x}_{0}^{+}\tilde{x}}\right)_{c_{1}d_{1}}\left(\tau^{j}\right)_{d_{1}e_{1}}\nn\\
&&\left(g_{\tilde{x}_{0}^{+}\tilde{x}}\right)_{c_{2}d_{2}}\left(\tau^{k}\right)_{d_{2}e_{2}}\left(g_{\tilde{x}_{0}^{+}\tilde{x}}\right)_{ab}\delta_{f_{1}h_{1}}\delta_{f_{2}h_{2}}\delta_{f_{3}h_{3}}.
\eea
Similarly, the following nonzero contributions can be found:\\
\vspace{0.2cm}

$\left(3\right)\left(3\right)\left(1\right)\left(3\right)\left(1\right)$, with the same choices of parameters as in (\ref{nonzero_b_1})
\bea
&&\int d\mu_{C}(\psi,\bar{\psi})\Bigl\{ \psi_{a_{1}\alpha_{1}f_{1}}(y)\bar{\psi}_{c_{1}\xi_{1}f}(\tilde{x}_{0}^{+})\psi_{a_{2}\alpha_{2}f_{2}}(y)\bar{\psi}_{c_{2}\xi_{2}h}(\tilde{x}_{0}^{+})\psi_{e_{1}\eta_{1}f}\left(\tilde{x}\right)\bar{\psi}_{b_{1}\beta_{1}h_{1}}\left(y_{\rho}^{\pi}\right)\nn\\
&&\psi_{b\beta F}\left(x_{\mu}^{\epsilon}\right)\bar{\psi}_{b_{2}\beta_{2}h_{2}}\left(y_{\rho}^{\pi}\right)\psi_{e_{2}\eta_{2}h}\left(\tilde{x}\right)\bar{\psi}_{b_{3}\beta_{3}h_{3}}\left(y_{\rho}^{\pi}\right)\psi_{a_{3}\alpha_{3}f_{3}}(y)\bar{\psi}_{a\alpha F}(x)\Bigr\}
\eea

\bea
&&-\frac{a}{8}\frac{\Xi^{6}}{Z_{f}}\epsilon_{c_{1}c_{2}a}\epsilon_{e_{1}e_{2}b}\left(1+\gamma_{0}\right)_{\alpha_{1}\beta_{1}}\left(1+\gamma_{0}\right)_{\alpha_{2}\beta_{3}}(1+\gamma_{0})_{\alpha_{3}\beta_{2}}\left(g_{\tilde{x}_{0}^{+}\tilde{x}}\right)_{c_{1}d_{1}}\left(\tau^{j}\right)_{d_{1}e_{1}}\nn\\
&&\left(g_{\tilde{x}_{0}^{+}\tilde{x}}\right)_{c_{2}d_{2}}\left(\tau^{k}\right)_{d_{2}e_{2}}\left(g_{\tilde{x}_{0}^{+}\tilde{x}}\right)_{ab}\delta_{f_{1}h_{1}}\delta_{f_{2}h_{3}}\delta_{f_{3}h_{2}}
\eea

$\left(3\right)\left(3\right)\left(2\right)\left(1\right)\left(2\right)$,

\bea
&&\int d\mu_{C}(\psi,\bar{\psi})\Bigl\{\psi_{a_{1}\alpha_{1}f_{1}}(y)\bar{\psi}_{c_{1}\xi_{1}f}(\tilde{x}_{0}^{+})\psi_{a_{2}\alpha_{2}f_{2}}(y)\bar{\psi}_{c_{2}\xi_{2}h}(\tilde{x}_{0}^{+})\psi_{e_{2}\eta_{2}h}\left(\tilde{x}\right)\bar{\psi}_{b_{1}\beta_{1}h_{1}}\left(y_{\rho}^{\pi}\right)\nn\\
&&\psi_{e_{1}\eta_{1}f}\left(\tilde{x}\right)\bar{\psi}_{b_{2}\beta_{2}h_{2}}\left(y_{\rho}^{\pi}\right)\psi_{b\beta F}\left(x_{\mu}^{\epsilon}\right)\bar{\psi}_{b_{3}\beta_{3}h_{3}}\left(y_{\rho}^{\pi}\right)\psi_{a_{3}\alpha_{3}f_{3}}(y)\bar{\psi}_{a\alpha F}(x)\Bigr\}
\eea

\bea
&&-\frac{a}{8}\frac{\Xi^{6}}{Z_{f}}\epsilon_{c_{1}c_{2}a}\epsilon_{e_{1}e_{2}b}\left(1+\gamma_{0}\right)_{\alpha_{1}\beta_{2}}\left(1+\gamma_{0}\right)_{\alpha_{2}\beta_{1}}(1+\gamma_{0})_{\alpha_{3}\beta_{3}}\left(g_{\tilde{x}_{0}^{+}\tilde{x}}\right)_{c_{1}d_{1}}\left(\tau^{j}\right)_{d_{1}e_{1}}\nn\\
&&\left(g_{\tilde{x}_{0}^{+}\tilde{x}}\right)_{c_{2}d_{2}}\left(\tau^{k}\right)_{d_{2}e_{2}}\left(g_{\tilde{x}_{0}^{+}\tilde{x}}\right)_{ab}\delta_{f_{1}h_{2}}\delta_{f_{2}h_{1}}\delta_{f_{3}h_{3}}
\eea

$\left(3\right)\left(3\right)\left(2\right)\left(3\right)\left(1\right)$,
\bea
&&-\int d\mu_{C}(\psi,\bar{\psi})\Bigl\{ \psi_{a_{1}\alpha_{1}f_{1}}(y)\bar{\psi}_{c_{1}\xi_{1}f}(\tilde{x}_{0}^{+})\psi_{a_{2}\alpha_{2}f_{2}}(y)\bar{\psi}_{c_{2}\xi_{2}h}(\tilde{x}_{0}^{+})\psi_{e_{2}\eta_{2}h}\left(\tilde{x}\right)\bar{\psi}_{b_{1}\beta_{1}h_{1}}\left(y_{\rho}^{\pi}\right)\nn\\
&&\psi_{b\beta F}\left(x_{\mu}^{\epsilon}\right)\bar{\psi}_{b_{2}\beta_{2}h_{2}}\left(y_{\rho}^{\pi}\right)\psi_{e_{1}\eta_{1}f}\left(\tilde{x}\right)\bar{\psi}_{b_{3}\beta_{3}h_{3}}\left(y_{\rho}^{\pi}\right)\psi_{a_{3}\alpha_{3}f_{3}}(y)\bar{\psi}_{a\alpha F}(x)\Bigr\} 
\eea

\bea    
&&-\frac{a}{8}\frac{\Xi^{6}}{Z_{f}}\epsilon_{c_{1}c_{2}a}\epsilon_{e_{1}e_{2}b}\left(1+\gamma_{0}\right)_{\alpha_{1}\beta_{3}}\left(1+\gamma_{0}\right)_{\alpha_{2}\beta_{1}}(1+\gamma_{0})_{\alpha_{3}\beta_{2}}\left(g_{\tilde{x}_{0}^{+}\tilde{x}}\right)_{c_{1}d_{1}}\left(\tau^{j}\right)_{d_{1}e_{1}}\nn\\
&&\left(g_{\tilde{x}_{0}^{+}\tilde{x}}\right)_{c_{2}d_{2}}\left(\tau^{k}\right)_{d_{2}e_{2}}\left(g_{\tilde{x}_{0}^{+}\tilde{x}}\right)_{ab}\delta_{f_{1}h_{3}}\delta_{f_{2}h_{1}}\delta_{f_{3}h_{2}}
\eea

$\left(3\right)\left(3\right)\left(4\right)\left(1\right)\left(1\right)$,

\bea
&&\int d\mu_{C}(\psi,\bar{\psi})\Bigl\{\psi_{a_{1}\alpha_{1}f_{1}}(y)\bar{\psi}_{c_{1}\xi_{1}f}(\tilde{x}_{0}^{+})\psi_{a_{2}\alpha_{2}f_{2}}(y)\bar{\psi}_{c_{2}\xi_{2}h}(\tilde{x}_{0}^{+})\psi_{b\beta F}\left(x_{\mu}^{\epsilon}\right)\bar{\psi}_{b_{1}\beta_{1}h_{1}}\left(y_{\rho}^{\pi}\right)\nn\\
&&\psi_{e_{1}\eta_{1}f}\left(\tilde{x}\right)\bar{\psi}_{b_{2}\beta_{2}h_{2}}\left(y_{\rho}^{\pi}\right)\psi_{e_{2}\eta_{2}h}\left(\tilde{x}\right)\bar{\psi}_{b_{3}\beta_{3}h_{3}}\left(y_{\rho}^{\pi}\right)\psi_{a_{3}\alpha_{3}f_{3}}(y)\bar{\psi}_{a\alpha F}(x)\Bigr\}
\eea

\bea
&&-\frac{a}{8}\frac{\Xi^{6}}{Z_{f}}\epsilon_{c_{1}c_{2}a}\epsilon_{e_{1}e_{2}b}\left(1+\gamma_{0}\right)_{\alpha_{1}\beta_{2}}\left(1+\gamma_{0}\right)_{\alpha_{2}\beta_{3}}(1+\gamma_{0})_{\alpha_{3}\beta_{1}}\left(g_{\tilde{x}_{0}^{+}\tilde{x}}\right)_{c_{1}d_{1}}\left(\tau^{j}\right)_{d_{1}e_{1}}\nn\\
&&\left(g_{\tilde{x}_{0}^{+}\tilde{x}}\right)_{c_{2}d_{2}}\left(\tau^{k}\right)_{d_{2}e_{2}}\left(g_{\tilde{x}_{0}^{+}\tilde{x}}\right)_{ab}\delta_{f_{1}h_{2}}\delta_{f_{2}h_{3}}\delta_{f_{3}h_{1}}
\eea

$\left(3\right)\left(3\right)\left(4\right)\left(2\right)\left(1\right)$,

\bea
&&-\int d\mu_{C}(\psi,\bar{\psi})\Bigl\{\psi_{a_{1}\alpha_{1}f_{1}}(y)\bar{\psi}_{c_{1}\xi_{1}f}(\tilde{x}_{0}^{+})\psi_{a_{2}\alpha_{2}f_{2}}(y)\bar{\psi}_{c_{2}\xi_{2}h}(\tilde{x}_{0}^{+})\psi_{b\beta F}\left(x_{\mu}^{\epsilon}\right)\bar{\psi}_{b_{1}\beta_{1}h_{1}}\left(y_{\rho}^{\pi}\right)\nn\\
&&\psi_{e_{2}\eta_{2}h}\left(\tilde{x}\right)\bar{\psi}_{b_{2}\beta_{2}h_{2}}\left(y_{\rho}^{\pi}\right)\psi_{e_{1}\eta_{1}f}\left(\tilde{x}\right)\bar{\psi}_{b_{3}\beta_{3}h_{3}}\left(y_{\rho}^{\pi}\right)\psi_{a_{3}\alpha_{3}f_{3}}(y)\bar{\psi}_{a\alpha F}(x)\Bigr\}
\eea
    
\bea \label{B-33421}
&&-\frac{a}{8}\frac{\Xi^{6}}{Z_{f}}\epsilon_{c_{1}c_{2}a}\epsilon_{e_{1}e_{2}b}\left(1+\gamma_{0}\right)_{\alpha_{1}\beta_{3}}\left(1+\gamma_{0}\right)_{\alpha_{2}\beta_{2}}(1+\gamma_{0})_{\alpha_{3}\beta_{1}}\left(g_{\tilde{x}_{0}^{+}\tilde{x}}\right)_{c_{1}d_{1}}\left(\tau^{j}\right)_{d_{1}e_{1}}\nn\\
&&\left(g_{\tilde{x}_{0}^{+}\tilde{x}}\right)_{c_{2}d_{2}}\left(\tau^{k}\right)_{d_{2}e_{2}}\left(g_{\tilde{x}_{0}^{+}\tilde{x}}\right)_{ab}\delta_{f_{1}h_{3}}\delta_{f_{2}h_{2}}\delta_{f_{3}h_{1}}
\eea

The rest of the terms are found to be related to the above terms in our computation as follows:

\[
\resizebox{0.95\hsize}{!}{$-\left(5\right)\left(3\right)\left(1\right)\left(1\right)\left(2\right)=-\left(5\right)\left(4\right)\left(1\right)\left(3\right)\left(1\right)=-\left(4\right)\left(3\right)\left(4\right)\left(2\right)\left(1\right)=-\left(4\right)\left(4\right)\left(2\right)\left(3\right)\left(1\right)=-\left(3\right)\left(3\right)\left(4\right)\left(1\right)\left(1\right)=-\left(3\right)\left(4\right)\left(2\right)\left(1\right)\left(2\right)
$},
\]
\[
\resizebox{0.95\hsize}{!}{$-\left(5\right)\left(3\right)\left(1\right)\left(3\right)\left(1\right)=-\left(5\right)\left(4\right)\left(1\right)\left(1\right)\left(2\right)=-\left(4\right)\left(3\right)\left(4\right)\left(1\right)\left(1\right)=-\left(4\right)\left(4\right)\left(2\right)\left(1\right)\left(2\right)=-\left(3\right)\left(3\right)\left(4\right)\left(2\right)\left(1\right)=-\left(3\right)\left(4\right)\left(2\right)\left(3\right)\left(1\right)$},
\]
\[
\resizebox{0.95\hsize}{!}{$-\left(5\right)\left(3\right)\left(2\right)\left(1\right)\left(2\right)=-\left(5\right)\left(4\right)\left(4\right)\left(1\right)\left(1\right)=-\left(4\right)\left(3\right)\left(2\right)\left(3\right)\left(1\right)=-\left(4\right)\left(4\right)\left(4\right)\left(2\right)\left(1\right)=-\left(3\right)\left(3\right)\left(1\right)\left(3\right)\left(1\right)=-\left(3\right)\left(4\right)\left(1\right)\left(1\right)\left(2\right)
$},
\]
\[
\resizebox{0.95\hsize}{!}{$-\left(5\right)\left(3\right)\left(2\right)\left(3\right)\left(1\right)=-\left(5\right)\left(4\right)\left(4\right)\left(2\right)\left(1\right)=-\left(4\right)\left(3\right)\left(2\right)\left(1\right)\left(2\right)=-\left(4\right)\left(4\right)\left(4\right)\left(1\right)\left(1\right)=-\left(3\right)\left(3\right)\left(1\right)\left(1\right)\left(2\right)=-\left(3\right)\left(4\right)\left(1\right)\left(3\right)\left(1\right)
$},
\]
\[
\resizebox{0.95\hsize}{!}{$-\left(5\right)\left(3\right)\left(4\right)\left(1\right)\left(1\right)=-\left(5\right)\left(4\right)\left(2\right)\left(1\right)\left(2\right)=-\left(4\right)\left(3\right)\left(1\right)\left(3\right)\left(1\right)=-\left(4\right)\left(4\right)\left(1\right)\left(1\right)\left(2\right)=-\left(3\right)\left(3\right)\left(2\right)\left(3\right)\left(1\right)=-\left(3\right)\left(4\right)\left(4\right)\left(2\right)\left(1\right)
$},
\]
\[
\resizebox{0.95\hsize}{!}{$-\left(5\right)\left(3\right)\left(4\right)\left(2\right)\left(1\right)=-\left(5\right)\left(4\right)\left(2\right)\left(3\right)\left(1\right)=-\left(4\right)\left(3\right)\left(1\right)\left(1\right)\left(2\right)=-\left(4\right)\left(4\right)\left(1\right)\left(3\right)\left(1\right)=-\left(3\right)\left(4\right)\left(4\right)\left(1\right)\left(1\right)=-\left(3\right)\left(3\right)\left(2\right)\left(1\right)\left(2\right)
$}.
\]
Here the overall sign appears with respect to $\epsilon_{c_{1}c_{2}a}\epsilon_{e_{1}e_{2}b}$, and we used $\Gamma^{-e_{0}}=-(1+\gamma_{0})$.

As for gauge-invariant composite fields, unlike the case of mesons, the explanation here is slightly more involved and requires some additional work. In particular, suppose that there are no $\epsilon$ symbols present in the baryon field vectors $F$ and $G$ (\ref{vect. baryons}), with $a_1 \neq a_2 \neq a_3$ and $b_1 \neq b_2 \neq b_3$. Then, upon examining any of the equations (\ref{B-33112})--(\ref{B-33421}), one sees that there is insufficient structure in the color indices to apply the determinant relation (\ref{det g}). As a result, the following gauge integral must be evaluated—e.g., in the case of (\ref{B-33112}). Notably, the gauge integral itself is identical across all terms (cf. (\ref{exp_bar})); only the contractions differ, leading to distinct results depending on the term in question.
\bea \label{gauge int.}
\int d\mu (g) \left(g_{\tilde{x}_{0}^{+}\tilde{x}}\right)_{c_{1}d_{1}}
\left(g_{\tilde{x}_{0}^{+}\tilde{x}}\right)_{c_{2}d_{2}}
\left(g_{\tilde{x}_{0}^{+}\tilde{x}}\right)_{ab},
\eea
performing the gauge integral (\ref{gauge int.}) using the approach discussed in \cite{Creutz:1978ub}, see also the integral $\mathcal{I}_6$ eq. (B4) in \cite{FariadaVeiga:2008zz} (note the misprint in the indices of the fourth and fifth terms), one finds, up to a numerical coefficient, the expression $\epsilon_{d_{1}d_{2}b}\epsilon_{c_{1}c_{2}a}$. Then, examining the contractions (\ref{F-contractions b1}) resulting from the Fermi integral (\ref{nonzero_b_1}), we find that the resulting term includes
\bea \label{gauge int. B1}
\epsilon_{d_{1}d_{2}b_3}\epsilon_{a_{1}a_{2}a_{3}}(\tau^j)_{d_1b_1}(\tau^k)_{d_2b_2}.
\eea
This expression simplifies immediately to
\bea
&\delta_{b_{3}a_{3}}(\tau^{j})_{a_{1}b_{1}}(\tau^{k})_{a_{2}b_{2}}
-\delta_{b_{3}a_{2}}(\tau^{j})_{a_{1}b_{1}}(\tau^{k})_{a_{3}b_{2}} -\delta_{b_{3}a_{3}}(\tau^{j})_{a_{2}b_{1}}(\tau^{k})_{a_{1}b_{2}}\nn\\
&+\delta_{b_{3}a_{1}}(\tau^{j})_{a_{2}b_{1}}(\tau^{k})_{a_{3}b_{2}}+\delta_{b_{3}a_{2}}(\tau^{j})_{a_{3}b_{1}}(\tau^{k})_{a_{1}b_{2}}-\delta_{b_{3}a_{1}}(\tau^{j})_{a_{3}b_{1}}(\tau^{k})_{a_{2}b_{2}}.
\eea
Clearly, for even a single nonzero contribution to arise, one of the conditions $b_3 = a_i $ for some $i = 1, 2, 3$ must be satisfied, in addition to the necessary conditions for producing the product $\tau^j \tau^k$  and taking the trace over it. However, if we contract the expression in (\ref{gauge int. B1}) with $\epsilon_{a_1 a_2 a_3} \epsilon_{b_1 b_2 b_3}$, as if the baryon field vectors in (\ref{vect. baryons}) were gauge-invariant, the result is generically nonzero, without requiring any prior constraint on the index $b_3$, and matches the expressions obtained above (after plugging in the correct numerical coefficient derived from (\ref{gauge int.})), see (\ref{tau^2}). The same logic applies to the remaining terms.

\section{Computational results}
\label{sec:appendixfinalresults}

Here we provide the computational results of the fermionic parts of the expectation values for baryons (\ref{E_f baryons}) and mesons (\ref{E_f mesons}). Recall we use the uppercase letters \(U\), \(D\), and \(S\) to denote what is commonly referred to as the up, down, and strange quarks, respectively.

Regarding the F-K-like formula (\ref{F-K}), special care is needed in handling computational results: while the LHS requires \(F = \Theta G\), the RHS (path integral) employs \(F\) and \(G\) as defined in (\ref{vect. mesons}) for mesons and (\ref{vect. baryons}) for baryons. Thus, when computing expectation values (\ref{E_f mesons}) and (\ref{E_f baryons}), the constraint 
\bea \label{F=ThG}
F = \Theta G
\eea
must be enforced.  
For mesons, given \(\alpha_i, \beta_i = +,-\) (where \(+\) denotes a quark and \(-\) an antiquark) and \(f_i, h_i = U, S, D\), we take \(F= \bar{\psi}_{+U} \psi_{-D}\). Then using (\ref{F=ThG}) by the anti-homomorphism property of \(\Theta\) (see below (\ref{Theta act.})), this implies \(G = \bar{\psi}_{-D} \psi_{+U}\). Such a restriction ensures expectation values correspond to physically meaningful composite states, e.g., the pion-like \(\pi^+\). Color indices, spacetime points, and time reflection are suppressed since they are redundant for numerical evaluations.  

For baryons, the situation is more intricate but follows the same principle: enforcing (\ref{F=ThG}), a configuration like \(\psi_{+U} \psi_{+U} \psi_{+D}\) yields \(G = \bar{\psi}_{+D} \bar{\psi}_{+U} \bar{\psi}_{+U}\), which aligns with a proton-like state.  

\begin{remark} \label{+-}
In 2-dimensional spacetime, the $\psi$ fields carry a natural double labeling of particle-antiparticle states (through bar and spin indices), since the spin index assumes only two values, + (particle) and - (antiparticle).
\end{remark}

\section*{Mesons}

\subsection*{Two-flavor case}

\begin{tabular}{|c|c|c|c|c|c|cc|c|c|c|c|c|c|}
\hline 
\multicolumn{1}{|c}{F} &  &  & \multicolumn{1}{c}{G} &  & $E_{f}^{\pm}$ &  &  & \multicolumn{1}{c}{F} &  &  & \multicolumn{1}{c}{G} &  & $E_{f}^{\pm}$\tabularnewline
\cline{1-2}\cline{4-6}\cline{9-10}\cline{12-14}
U & U &  & U & U & 4 &  &  & D & D &  & D & D & 4\tabularnewline
\cline{1-2}\cline{4-5}\cline{9-10}\cline{12-13}
D & U &  & U & D &  &  &  & U & D &  & D & U & \tabularnewline
\cline{1-2}\cline{4-6}\cline{9-10}\cline{12-14}
\end{tabular}\\

\vspace{1cm}

\subsection*{Three-flavor case}
\begin{tabular}{|c|c|c|c|c|c|cc|c|c|c|c|c|c|cc|c|c|c|c|c|c|}
\hline 
\multicolumn{1}{|c}{F} &  &  & \multicolumn{1}{c}{G} &  & $E_{f}^{\pm}$ &  &  & \multicolumn{1}{c}{F} &  &  & \multicolumn{1}{c}{G} &  & $E_{f}^{\pm}$ &  &  & \multicolumn{1}{c}{F} &  &  & \multicolumn{1}{c}{G} &  & $E_{f}^{\pm}$\tabularnewline
\cline{1-2}\cline{4-6}\cline{9-10}\cline{12-14}\cline{17-18}\cline{20-22}
U & U &  & U & U &  &  &  & D & D &  & D & D &  &  &  & S & S &  & S & S & \tabularnewline
\cline{1-2}\cline{4-5}\cline{9-10}\cline{12-13}\cline{17-18}\cline{20-21}
D & U &  & U & D & 4 &  &  & U & D &  & D & U & 4 &  &  & U & S &  & S & U & 4\tabularnewline
\cline{1-2}\cline{4-5}\cline{9-10}\cline{12-13}\cline{17-18}\cline{20-21}
S & U &  & U & S &  &  &  & S & D &  & D & S &  &  &  & D & S &  & S & D & \tabularnewline
\cline{1-2}\cline{4-6}\cline{9-10}\cline{12-14}\cline{17-18}\cline{20-22}
\end{tabular}

\vspace{2cm}

\section*{Baryons}

\subsection*{Three-flavor case}

\resizebox{0.9\hsize}{!}{\begin{tabular}{|c|c|c|c|c|c|c|c|cc|c|c|c|c|c|c|c|c|cc|c|c|c|c|c|c|c|c|}
\hline 
\multicolumn{1}{|c}{} & \multicolumn{1}{c}{F} &  &  & \multicolumn{1}{c}{} & \multicolumn{1}{c}{G} &  & $E_{f}^{\pm}$ &  &  & \multicolumn{1}{c}{} & \multicolumn{1}{c}{F} &  &  & \multicolumn{1}{c}{} & \multicolumn{1}{c}{G} &  & $E_{f}^{\pm}$ &  &  & \multicolumn{1}{c}{} & \multicolumn{1}{c}{F} &  &  & \multicolumn{1}{c}{} & \multicolumn{1}{c}{G} &  & $E_{f}^{\pm}$\tabularnewline
\cline{1-3}\cline{5-8}\cline{11-13}\cline{15-18}\cline{21-23}\cline{25-28}
U & U & U &  & U & U & U & 48 &  &  & D & D & D &  & D & D & D & 48 &  &  & S & S & S &  & S & S & S & 48\tabularnewline
\cline{1-3}\cline{5-8}\cline{11-13}\cline{15-18}\cline{21-23}\cline{25-28}
D & U & U &  & U & U & D &  &  &  & U & D & D &  & D & D & U &  &  &  & U & S & S &  & S & S & U & \tabularnewline
\cline{1-3}\cline{5-7}\cline{11-13}\cline{15-17}\cline{21-23}\cline{25-27}
U & D & U &  & U & D & U &  &  &  & D & U & D &  & D & U & D &  &  &  & S & U & S &  & S & U & S & \tabularnewline
\cline{1-3}\cline{5-7}\cline{11-13}\cline{15-17}\cline{21-23}\cline{25-27}
U & U & D &  & D & U & U &  &  &  & D & D & U &  & U & D & D &  &  &  & S & S & U &  & U & S & S & \tabularnewline
\cline{1-3}\cline{5-7}\cline{11-13}\cline{15-17}\cline{21-23}\cline{25-27}
S & U & U &  & U & U & S & 16 &  &  & S & D & D &  & D & D & S & 16 &  &  & D & S & S &  & S & S & D & 16\tabularnewline
\cline{1-3}\cline{5-7}\cline{11-13}\cline{15-17}\cline{21-23}\cline{25-27}
U & S & U &  & U & S & U &  &  &  & D & S & D &  & D & S & D &  &  &  & S & D & S &  & S & D & S & \tabularnewline
\cline{1-3}\cline{5-7}\cline{11-13}\cline{15-17}\cline{21-23}\cline{25-27}
U & U & S &  & S & U & U &  &  &  & D & D & S &  & S & D & D &  &  &  & S & S & D &  & D & S & S & \tabularnewline
\cline{1-3}\cline{5-8}\cline{11-13}\cline{15-18}\cline{21-23}\cline{25-28}
U & D & S &  & S & D & U & 8 &  &  & D & U & S &  & S & U & D & 8 &  &  & S & U & D &  & D & U & S & 8\tabularnewline
\cline{1-3}\cline{5-7}\cline{11-17}\cline{21-27}
U & S & D &  & D & S & U &  &  &  & D & S & U &  & U & S & D &  &  &  & S & D & U &  & U & D & S & \tabularnewline
\cline{1-3}\cline{5-8}\cline{11-13}\cline{15-18}\cline{21-23}\cline{25-28}
\end{tabular}}

\subsection*{Two-flavor case}
\resizebox{0.8\hsize}{!}{\begin{tabular}{|c|c|c|c|c|c|c|c|cc|c|c|c|c|c|c|c|c|}
\hline 
\multicolumn{1}{|c}{} & \multicolumn{1}{c}{F} &  &  & \multicolumn{1}{c}{} & \multicolumn{1}{c}{G} &  & $E_{f}^{\pm}$ &  &  & \multicolumn{1}{c}{} & \multicolumn{1}{c}{F} &  &  & \multicolumn{1}{c}{} & \multicolumn{1}{c}{G} &  & $E_{f}^{\pm}$\tabularnewline
\cline{1-3}\cline{5-8}\cline{11-13}\cline{15-18}
U & U & U &  & U & U & U & 48 &  &  & D & D & D &  & D & D & D & 48\tabularnewline
\cline{1-3}\cline{5-8}\cline{11-13}\cline{15-18}
D & U & U &  & U & U & D &  &  &  & U & D & D &  & D & D & U & \tabularnewline
\cline{1-3}\cline{5-7}\cline{11-13}\cline{15-17}
U & D & U &  & U & D & U & 16 &  &  & D & U & D &  & D & U & D & 16\tabularnewline
\cline{1-3}\cline{5-7}\cline{11-13}\cline{15-17}
U & U & D &  & D & U & U &  &  &  & D & D & U &  & U & D & D & \tabularnewline
\cline{1-3}\cline{5-8}\cline{11-18}
\end{tabular}}\\

While the above values are computed for baryons, with all $+$'s, antibaryons give the same values as above with $+\to-$ and $y=\tilde{x}$.

\section{The naive continuum limit: covariantly conserved matter current} \label{appndx. cont. lim.}

Taking the finite difference of the coefficient (\ref{lat_current}) at $x$ and $x^-_\mu$ gives
\bea
&&\frac{1}{a}\left(J_{\mu}^{j}(x)-J_{\mu}^{j}(x_{\mu}^{-})\right)\nn\\
      &&=\sum_{\mu}\frac{1}{2a}\left(\bar{\psi}(x_{\mu}^{+})+\bar{\psi}(x_{\mu}^{-})\right)\tau^{j}\psi\left(x\right)-\sum_{\mu}\frac{1}{2a}\left(\bar{\psi}\left(x_{\mu}^{+}\right)+\bar{\psi}\left(x_{\mu}^{-}\right)\right)\tau^{j}\psi(x)\nn\\
	&&+\sum_{\mu}\frac{1}{2a}\left(\bar{\psi}(x_{\mu}^{+})-\bar{\psi}(x_{\mu}^{-})\right)\gamma_{\mu}\tau^{j}\psi\left(x\right)+\sum_{\mu}\frac{1}{2a}\bar{\psi}(x)\gamma_{\mu}\tau^{j}\left(\psi\left(x_{\mu}^{+}\right)-\psi\left(x_{\mu}^{-}\right)\right)\nn\\
&&+\sum_{\mu}\frac{iag}{2a}\left(\bar{\psi}(x_{\mu}^{+})A_{\mu}(x_{\mu}^{+})-\bar{\psi}(x_{\mu}^{-})A_{\mu}(x)\right)\tau^{j}\psi\left(x\right)\nn\\
	&&+\sum_{\mu}\frac{iag}{2a}\bar{\psi}(x)\tau^{j}\left(A_{\mu}(x_{\mu}^{+})\psi\left(x_{\mu}^{+}\right)-A_{\mu}(x)\psi\left(x_{\mu}^{-}\right)\right)\nn\\
	&&+\sum_{\mu}\frac{iag}{2a}\left(\bar{\psi}(x_{\mu}^{+})\gamma_{\mu}A_{\mu}(x_{\mu}^{+})+\bar{\psi}(x_{\mu}^{-})\gamma_{\mu}A_{\mu}(x)\right)\tau^{j}\psi\left(x\right)\nn\\
	&&-\sum_{\mu}\frac{iag}{2a}\bar{\psi}(x)\gamma_{\mu}\tau^{j}\left(A_{\mu}(x_{\mu}^{+})\psi\left(x_{\mu}^{+}\right)+A_{\mu}(x)\psi\left(x_{\mu}^{-}\right)\right),
\eea
where we transformed the time translation of $x$ in the second term from $\psi$ to $\bar{\psi}$. This change of variables should be understood in the context of translationally invariant expectation values, with a slight abuse of notation. Nonetheless, in principle, the derivation above should be carried out more systematically to rigorously obtain the Slavnov–Taylor identity, analogous to the Ward–Takahashi identity for the abelian gauge group (see \cite{Montvay:1994cy}). Then using the finite difference derivative of the variables $\psi(x)$'s and $\bar{\psi}(x)$'s
\bea
\partial_{\mu}^{a}\psi(x)\doteq\frac{1}{2a}\left(\psi\left(x_{\mu}^{+}\right)-\psi\left(x_{\mu}^{-}\right)\right)
\eea
and keeping the leading order terms, in particular, for
\bea
\psi(x_{\mu}^{\pm})=\psi(x)\pm\mathcal{O}(a)
\eea
and
\bea
A_{\mu}(x_{\mu}^{\pm})=A(x)\pm\mathcal{O}(a),
\eea
we find the \emph{naive} continuum limit as follows.
\bea
&\lim\limits_{a \to 0}\frac{1}{a}\left(J_{\mu}^{j}(x)-J_{\mu}^{j}(x_{\mu}^{-})\right)\nn\\
&=\sum_{\mu}\left(\partial_{\mu}+ig\left[A_{\mu}(x),\cdot\,\right]\right)\bar{\psi}(x)\gamma_{\mu}\tau^{j}\psi\left(x\right)=\sum_{\mu}D_{\mu}\left(\bar{\psi}(x)\gamma_{\mu}\tau^{j}\psi\left(x\right)\right).
\eea
Since the limit yields the non-abelian \emph{classical} current with the covariant derivative denoted by $D_\mu$, we conclude the current is conserved, see e.g. \cite{Weinberg:1996kr}.

\end{document}